\documentclass[12pt]{article}
\usepackage{graphicx}
\begin{document}

\title{In medium $T$-matrix for superfluid nuclear matter}
\baselineskip=1. \baselineskip

\author{     P. Bo\.{z}ek\thanks{electronic
address~
bozek@solaris.ifj.edu.pl}\\
  Institute of Nuclear Physics, PL-31-342 Krak\'{o}w, Poland}
\date{\today}

\maketitle

\vskip .3cm


\vskip .3cm

\begin{abstract}
We study a generalized ladder resummation in the superfluid phase
of the nuclear matter. 
The approach
is based on a conserving generalization of the usual $T$-matrix approximation
 including also anomalous self-energies and propagators.  
The approximation here
discussed is  a 
 generalization of the  usual mean-field BCS approach and of the in medium 
$T$-matrix approximation in the normal phase.
The numerical 
results in this
work are obtained in the quasi-particle approximation.
Properties of the resulting self-energy,
superfluid gap and spectral functions are studied. 
\end{abstract}

\vskip .4cm

{\bf 21.65+f, 24.10Cn, 26.60+c}

\vskip .4cm

\section{Introduction}

One of the most general properties of fermion systems with 
attractive interactions is the transition to a superfluid state at finite
density and low temperature. It is expected that such a phenomenom happens
also for the strongly interacting system of nucleons, {\it i.e} for 
cold nuclear matter. 
Calculations based on nucleon-nucleon interaction predict very large values of 
the superfluid gap. Typically around $5-10$MeV for the isospin singlet ($T=0$)
partial-wave $^3S_1-^3D_1$ \cite{sd1,sd2,sd3,sdd}. The actual value of the
superfluid gap is a matter of debate, because of expected screening and
self-energy corrections. One has to note that also some of the phenomenological
potentials fitted to the pairing properties of finite nuclei give significant
values of the superfluid gap in nuclear matter \cite{lda}.
 The study of $T=0$ pairing in the systematics of 
$N\simeq Z$ nuclei became possible with advent of radioactive beam facilities.
This lead to a resurgence of the study of the nuclear mass systematics 
\cite{g1,rpair,sa}. Thus it is of importance to obtain results on the
nature of the nuclear pairing ($T=0$ or $T=1$) and the value of the superfluid
gap for symmetric nuclear matter at saturation density and below.
It appears that the mean-field gap equation without medium modifications is
unrealistic \cite{cl,wam,screen,ja3,seb,ses}
 and hence the best strategy would be to calculate a
 density dependent gap for nuclear matter (including relevant many-body
 corrections) and   use it 
 in a local density approximation  for 
 calculations in finite systems \cite{lda,rpair}.

The neutron rich nuclear matter is used in modeling of 
the crust and of the core of
neutron stars. It is generally believed that such an asymmetric nuclear matter
is superfluid, with different kinds of superfluid gap appearing in the vast
range of densities present in the neutron star. The value of the
superfluid gap is of importance for  fundamental problems in the 
neutron stars, the formation of glitches, the value of the viscosity and 
 the cooling rates in different 
scenarios \cite{tsu}.  Again one of the possible approaches is to calculate the
superfluid gap from  bare $NN$ interaction using the Brueckner-Hartree-Fock 
 approximation to get single-particle energies for the gap equation
\cite{nms1,nms2,screen,nms3,nms4,seb,ses,ro3}. However, 
in order to obtain reliable
estimates for the superfluid gap in the neutron matter we need to have under
control  in medium many-body corrections to the gap equation. 

As noted above strong modifications of the mean-field
 result for the superfluid
gap
are expected in the nuclear matter due to screening effects 
\cite{cl,wam,screen},
modifications of the effective mass and self-energy corrections
 \cite{ja3,ses,seb}. One of the motivations of the present study is to
 investigate
another
source of in medium correction to the value of the gap. These corrections
occur in a generalization of the resummation of ladder diagrams to include
also anomalous propagators. In the following we study the formalism and give
numerical estimates of the corrections to the anomalous self-energy.
These additional terms introduce an energy dependence in the superfluid gap,
but the modifications of the value of the gap at the Fermi surface are not as
dramatic as from the other medium effects.


The correction to the binding energy due to the superfluid 
rearrangement of the ground
state is believed to be small \cite{walecka}.
 However, some of the calculations using
realistic nuclear forces predict quite large values of the superfluid gap 
in the nuclear matter. A large superfluid gap could lead to modifications of
the normal part of the self-energy and the spectral function. In the following
we study a consistent approximation treating on equal footing the normal and
anomalous part of the self-energy. We find that important  modifications of
the single-particle and two-body propagators appear if the superfluid gap is 
large. These significant modifications  in the
superfluid  present, {\it a posteriori}, an important reason to consider the
generalized formalism discussed in this work. 
 Also the expansion of the ground state energy or other quantities around
the wrong ground state is not satisfactory for a theory aiming at the
description of the many-body problem from first-principles, using free
nucleon-nucleon potentials. The incorrect ground state could  also lead
to instabilities in the actual iterative numerical solution of
 the many-body equations, related to the appearance of the Cooper
instability \cite{cooper}.

Mean-field approaches give a qualitatively correct
 description of the formation of the
superfluid gap by the BCS mechanism, but fail  in the resummation of the
hard core. Recently superfluid nuclear matter
 was studied in an approach starting from the
in medium $T$-matrix approximation \cite{ja2}. 
The resummation of  ladder
diagrams in the $T$-matrix can be used to deal with the hard core in the
interaction potential.
The $T$-matrix approximation   for the self-energy was studied intensively
 in the last decade
 in normal nuclear matter. 
It is different from the usual
$G$-matrix approximation.
 The $T$-matrix formalism,  also
called self-consistent Green's function approach, can be used directly
 at high temperature, above
the superfluid phase transition\cite{dickhoff1,dickhoff2,ja,ja4}.
It allows to study in a self-consistent way the one-particle self-energies 
 and spectral functions \cite{dickhoff1,ja,gent,rs,v1},
 the two-particle properties
 and in medium cross-sections \cite{dickhoff1,dickhoff2,ja,ja4,jaconf,rs},
the onset of superfluidity \cite{roepke,ro2,ro3,ja2,rs},
 the self-energy corrections to
the superfluid gap \cite{ja3}. The treatment of 
 the old question of saturation properties
of nuclear matter in the $T$-matrix approximation is at the present stage  not
superior to the most recent $G$ matrix or variational 
calculations, including realistic
interactions, three-body and three-body forces corrections. However, the
$T$-matrix self-energy leads to reliable results for the single particle 
properties, in particular it gives a consistent value for the Fermi energy
fulfilling the Hugenholz-Van Hove theorem \cite{hvh,ja4}.

Besides the above motivations to develop the $T$-matrix approach for nuclear
 matter, this approximation seems to be the most natural starting point 
for the study of the
 superfluid phase of the nuclear matter.
 The appearance of a singularity in the
$T$-matrix at zero total momentum of the pair and at twice the Fermi energy
signals the formation of a long range order and defines the Thouless criterion
for the critical temperature. It is in fact equivalent to the critical
temperature corresponding to the appearance of  a nontrivial 
solution of the BCS gap equation. The same is true if off-shell
propagators are used in the $T$-matrix ladders. It corresponds to a
generalization of the gap equation to one using full spectral functions,
including the imaginary part of the normal self-energy 
\cite{ja3}. The study of the singularity in the $T$-matrix equations at finite
 temperature, in the normal phase, allows to identify the critical temperature
 and the precritical modifications of the spectral function, in medium
 cross-sections and density of states 
\cite{roepke,ro2,ro3,ro4,rs,ja,ja2,Schnell}.

The generalization of the $T$-matrix approximation to the
superfluid phase appearing {\it below} $T_c$ was discussed in Ref \cite{ja2}.
The approach is based on the observation that
 the formation of the superfluid order parameter requires a long
range order. This long range order, representing the formation of Cooper pairs 
at zero momentum and twice the Fermi energy,  corresponds to a singularity in
the $T$-matrix for the same energy and momentum as in the Thouless criterion.
Thus also {\it below} $T_c$ we expect the singularity in the two-body
propagator at  twice the Fermi energy.
The kernel of the $T$-matrix equation is modified in the superfluid
so that the singularity of the $T$-matrix equation is again equivalent to 
the BCS gap equation for non-zero values of the order parameter \cite{ja2}.

In the present work we investigate a different approach to the $T$-matrix
resummation in the superfluid. It is a generalization of the 
ordinary $T$-matrix ladder diagrams which includes also anomalous propagators.
Thus  normal and anomalous self-energies are calculated in a unified way.
 The approximation deals at the same time with ladder diagrams
 resummation for
the self-energy and  with the appearance of the order parameter.
The properties of the generalized $T$-matrix, self-energy and spectral
function  are discussed.
 A justification of the heuristic procedure of
Ref. \cite{ja2} is then obtained if the superfluid order parameter is
restricted to the BCS contribution.
 The practical calculations are more difficult 
in the generalized scheme here discussed. The number of propagators and
$T$-matrix components is doubled because of the appearance of the off-diagonal,
anomalous propagators. The numerical results presented in this work are
obtained in the
quasi-particle approximation, starting from mean-field BCS propagators.

\section{Green's functions in the superfluid}

\subsection{Notation and formulas for the normal phase}
We consider  infinite homogenous nuclear matter interacting through a
two-body potential. The energies are defined with respect to the 
chemical potential $\mu$
\begin{eqnarray}
H=\sum_\alpha \int d^3 x \ \Psi_\alpha^\dagger(x)(-\frac{\Delta}{2 m}-\mu)
\Psi_\alpha(x) \nonumber \\
 + \sum_{\alpha^{'},\beta^{'},\alpha,\beta}
\frac{1}{2}\int d^3 x \int d^3 y \Psi_{\alpha^{'}}^\dagger(x) 
\Psi_{\beta^{'}}^\dagger(y)  V_{\alpha^{'},\beta^{'},\alpha,\beta}(x,y)
\Psi_{\beta}(y) 
\Psi_\alpha(x) \ .
\end{eqnarray}
In the real time formalism the Green's functions are defined on a contour
in the time plane \cite{keldysh}. It means that as  functions of a single
valued time variable the Green's functions acquire a matrix structure
\begin{equation}
\label{ctime}
{\hat{G}_{\alpha\beta}(x_1,t_1;x_2,t_2)} = \left( \matrix{ G_{\alpha\beta}^{c}
(x_1,t_1;x_2,t_2)
& G_{\alpha\beta}^{<}(x_1,t_1;x_2,t_2) \cr
 G_{\alpha\beta}^{>}(x_1,t_1;x_2,t_2)
& G_{\alpha\beta}^{a}(x_1,t_1;x_2,t_2) }\right) \ \ \ .
\end{equation}
Where $G_{\alpha\beta}^{c}(x_1,t_1;x_2,t_2)= i   \langle T\Psi_\alpha(x_1,t_1) 
\Psi_\beta^\dagger(x_2,t_2) \rangle$ 
is the chronological Green's function, $G^{a}$ is
the anti-chronological Green's function and
 $G_{\alpha\beta}^{<}(x_1,t_1;x_2,t_2)= i \langle \Psi_\alpha(x_1,t_1)
 \Psi_\beta^\dagger(x_2,t_2) \rangle$
and $G_{\alpha\beta}^{<}(x_1,t_1;x_2,t_2)= -i\langle \Psi_\beta^\dagger(x_2,t_2)
\Psi_\alpha(x_1,t_1) \rangle$
are the correlation functions.
In a homogeneous system,  after Fourier transforming in the relative coordinate
and time, single-particle Green's functions depend 
on a single energy and momentum. Noting that the 
Green's function are diagonal in the spin-isospin indices, we can write 
\begin{equation}
\hat{G}_{\alpha\beta}(p,\omega)=\delta_{\alpha\beta}
\hat{G}(p,\omega) \ .
\end{equation}
A single scalar function (in spin-isospin indices) 
$G$ is sufficient to describe a spin and isospin
symmetric nuclear matter. 
The scalar  correlation functions are written using the spectral function
\begin{eqnarray}
G^{<}(p,\omega)=i f(\omega) A(p,\omega) \\
G^{>}(p,\omega)=-i(1-f(\omega)) A(p,\omega) \ \ .
\end{eqnarray}
$f(\omega)=\frac{1}{1+\exp{-\beta(\omega-\mu)}}$
is the Fermi distribution at temperature $T=1/\beta$.
The spectral function $A$ can be related to the retarded Green's function
\begin{eqnarray}
A(p,\omega)&=&-2 {\rm Im} G^{+}(p,\omega) \nonumber \\
& =& i\big( G^{>}(p,\omega)-G^{<}(p,\omega)\big)   \\
&=&\frac{-2{\rm Im}\Sigma^{+}(p,\omega)}{
(\omega-p^2/2 m-{\rm Re} \Sigma^+(p,\omega)+\mu)^2+{\rm Im}
\Sigma^{+}(p,\omega)^2 } \ ,
\label{normalspec}
\end{eqnarray}
where $\Sigma^{+}(p,\omega)$ is the retarded self-energy.

The $T$-matrix equation in the normal phase is 
\begin{eqnarray}
\label{teq} 
\langle {\bf p}|T_{\alpha^{'}\beta^{'}\alpha\beta}
^\pm({\bf P},\omega)|{\bf p}^{'}\rangle& =& V_{\alpha^{'}\beta^{'}\alpha\beta}
({\bf p},{\bf p}^{'}) \nonumber \\ & & + \sum_{\gamma \delta}
 \int\frac{d^3k}{(2 \pi)^3}
\int\frac{d^3q}{(2 \pi)^3} V_{\alpha^{'}\beta^{'}\gamma\delta}
({\bf p},{\bf k}) \nonumber \\ 
& &\langle {\bf k}|{\cal G}^\pm({\bf P},\omega)|{\bf q}\rangle
 \langle {\bf q}|T_{\gamma\delta\alpha\beta}^\pm({\bf P},\omega)
|{\bf p}^{'}\rangle  \ ,
\end{eqnarray}
where ${\cal G}$ is the disconnected retarded two-particle propagator 
\begin{eqnarray}
\label{twpro}
\langle {\bf p}|{\cal G}^\pm({\bf P},\omega)|{\bf p}^{'}\rangle  =  
(2 \pi)^3 \delta^3({\bf p}-{\bf p}^{'})\int \frac{d\omega^{'}}{2 \pi}
\int \frac{d\omega^{''}}{2 \pi} \nonumber \\
\Big( G^<({\bf P}/2+{\bf p},\omega^{''}-\omega^{'})G^<({\bf P}/2-{\bf p},
\omega^{'}) \nonumber \\ -G^>({\bf P}/2+{\bf p},\omega^{''}-
\omega^{'})G^>({\bf P}/2-{\bf p},\omega^{'}) \Big)/ 
\Big(\omega -\omega^{''} \pm 
i\epsilon\Big)  \nonumber \\
=(2 \pi)^3 \delta^3({\bf p}-{\bf p^{'}})
 {\cal G}^\pm({\bf P},\omega,{\bf p}) \ . 
\end{eqnarray}
Note that it is sufficient to solve a single equation for the retarded
$T$-matrix instead of a matrix equation in the indices on the time contour in
the complex time plane (\ref{ctime}).

A full structure in the spin-isospin indices and relative angles of momenta
must be kept. However,
usually a partial wave expansion of the $T$-matrix equation
is performed 
\begin{eqnarray}
\label{teqp}
\langle {p}|T^{(JST) \ \pm}_{l^{'}l}
({ P},\omega)|{ p}^{'}\rangle & =& V^{(JST)}_{l^{'}l}
({ p},{ p}^{'})+ \sum_{l^{''}}
 \int\frac{k^2 d k}{2 \pi^2}
 V_{l^{'}l^{''}}^{(JST)}
({ p},{ k}) \nonumber \\ 
& &{\cal G}^\pm({ P},\omega,k)
 \langle { k}|T^{(JST) \ \pm}_{l^{''}l}({ P},\omega)
|{ p}^{'}\rangle  \ 
\end{eqnarray}
after angle averaging the 
Kernel ${\cal G}$ 
\begin{equation}
{\cal G}^\pm( P,\omega, p)=\int\frac{d\Omega}{4\pi} 
{\cal G}^\pm({\bf P},\omega,{\bf p}) \ .
\end{equation}

\subsection{Anomalous Green's function}

In the superfluid phase the ground state of the system has a nonzero order
parameter, corresponding to bound Cooper pairs \cite{schrieffer}. 
This leads to new Green's functions
\begin{equation}
\label{anctime}
{\hat{F}_{\alpha\beta}(x_1,t_1;x_2,t_2)} = \left( \matrix{ F_{\alpha\beta}^{c}(x_1,t_1;x_2,t_2)
& F_{\alpha\beta}^{<}(x_1,t_1;x_2,t_2) \cr
 F_{\alpha\beta}^{>}(x_1,t_1;x_2,t_2)
& F_{\alpha\beta}^{a}(x_1,t_1;x_2,t_2) }\right) \ \ \ .
\end{equation}
$F^c$ is  the time ordered anomalous Green's function
$$F_{\alpha\beta}^{c}(x_1,t_1;,x_2,t_2)= i \langle T(\Psi_\alpha(x_1,t_1)
\Psi_\beta(x_2,t_2))\rangle $$ and $F^{a}$ the anti-chronological one
and where
\begin{eqnarray}
F_{\alpha\beta}^{<}(x_1,t_1;,x_2,t_2)&= &i \langle \Psi_\alpha(x_1,t_1)
\Psi_\beta(x_2,t_2)\rangle  \noindent \\
F_{\alpha\beta}^{>}(x_1,t_1;,x_2,t_2)&= &- i \langle \Psi_\beta(x_2,t_2)
\Psi_\alpha(x_1,t_1)\rangle  \ \ .
\end{eqnarray}
We consider a homogeneous infinite system so that the anomalous 
Green's functions depend on a single momentum and energy. 
Analogously as for the normal Green's functions we can write
\begin{eqnarray}
F_{\alpha\beta}^{<}(p,\omega)&= &i f(\omega) B_{\alpha\beta}
(p,\omega) \noindent \\
F_{\alpha\beta}^{>}(p,\omega)&= &-i (1-f(\omega)) B_{\alpha\beta}
(p,\omega)   \ \ . 
\end{eqnarray}
The anomalous spectral function is related to the imaginary part of the 
retarded propagator
\begin{eqnarray}
B_{\alpha\beta}(p,\omega)&=&-{\rm Im} \big(F^{+}_{\alpha\beta}
(p,\omega+i\epsilon\big)
-F^{+}_{\alpha\beta}(p,\omega-i\epsilon))\noindent \\
&=&i\big(F^{>}_{\alpha\beta}(p,\omega)
-F^{<}_{\alpha\beta}(p,\omega)\big)
\end{eqnarray}
It should be noted that the spectral function for the diagonal part 
of the Green's function is modified in the presence of the off-diagonal
self-energy $\Delta$.
We shall denote it by 
\begin{equation}
A_s(p,\omega)=-2{\rm Im} G^{+}(p,\omega)
\end{equation}
and reserve the notation $A(p,\omega)$ to the spectral function obtained by
putting $\Delta=0$ (Eq. \ref{normalspec}).
 
The spin-isospin structure of the anomalous Green's function is assumed to be
of the spin (isospin) singlet or triplet kind.
We write
\begin{equation}
\label{find}
F_{\alpha \beta}(p,\omega)= \tilde{\Delta}_{\alpha \beta} F(p,\omega) \ \ .
\end{equation}
In general the matrix $\tilde{\Delta}_{\alpha\beta}$ 
could depend on the momentum, 
energy and relative directions of spin, isospin and momentum.
To simplify we use in the following angle-averaged double propagators in the
$T$-matrix ladder.
The matrix $\tilde{\Delta}$ in spin-isospin fulfills 
\begin{equation}
\tilde{\Delta}^{\dagger}_{\alpha \beta}\tilde{\Delta}_{\beta \gamma}
=|D|^2 \delta_{\alpha\gamma} 
\end{equation}
for time-reversal invariant states.
Without loss of generality we can put $|D|^2=1$.
More specifically the spin (isospin) part of $\tilde{\Delta}_{\alpha\beta}$
is
\begin{equation}
\left( \matrix{ d_x +i d_y& -i d_z \cr 
-i d_z & d_x-i d_y} \right)
\end{equation}
for the triplet gap ($d_x^2+d_y^2+d_z^2=1$) and
\begin{equation}
\left( \matrix{ 0 & i  \cr 
-i  & 0 } \right)
\end{equation}
for the singlet one.
Together with the choice of a diagonal normal Green's function we describe
the propagators using  scalar functions in spin-isospin indices
$F^{<>}(p,\omega)$ and $G^{<>}(p,\omega)$ or the corresponding retarded 
propagators and spectral functions.


\subsection{Ladder resummation in the superfluid}
\label{superladder}

 Ladder resummations in the superfluid have been considered in the description
 of high $T_c$ superconductors \cite{haussman}.
A thermodynamically consistent scheme which reduces to the usuall
$T$-matrix equation above $T_c$ can be constructed. The simplest way is to
 introduce a generalised $T$-matrix with aditional indices indicating 
if the incoming line is anomaleous or normal, following the Nambu formalism
 for superconductors. Rescticting oneself to
two normal or two anomalous propagators in the ladder leads to the following
 additional 
matrix structure in the $T$-matrix
\begin{eqnarray}
\label{eqtl}
\left( \matrix{ \langle {\bf p}|T^{\pm}_{\alpha^{'}\beta^{'}\alpha \beta}({\bf P},
\Omega)|{\bf p^{'}}\rangle  & \langle {\bf p}|L^{\pm}_{\alpha^{'}\beta^{'}\alpha \beta}
({\bf P},
\Omega)|{\bf p^{'}}\rangle  \cr
 \langle {\bf p}|L^{\pm \  \dagger}_{\alpha^{'}\beta^{'}\alpha \beta}({\bf P},
-\Omega)|{\bf p^{'}}\rangle  &
 \langle {\bf p}|T^{\pm \dagger}_{\alpha^{'}\beta^{'}\alpha \beta}({\bf P},
-\Omega)|{\bf p^{'}}\rangle  
 } \right) = \nonumber \\
\left( \matrix{ V({\bf p},{\bf p^{'}})_{\alpha^{'}\beta^{'}\alpha \beta}
& 0 \cr  0& V({\bf p},{\bf p^{'}})_{\alpha^{'}\beta^{'}\alpha \beta}
 } \right) 
+ \int \frac{d^3q}{(2 \pi)^3}
 \sum_{\gamma \delta}
\left( \matrix{ V({\bf p},{\bf q})_{\alpha^{'}\beta^{'}\gamma^{'} \delta^{'}}
& 0 \cr 0 &  V({\bf p},{\bf q})_{\alpha^{'}\beta^{'}\gamma^{'} \delta^{'}}
 } \right)\nonumber \\
 \left( \matrix{ \delta_{\gamma \gamma^{'} }\delta_{\delta \delta^{'}  }
{\cal G}^\pm({ \bf P},\Omega,{\bf q}) &
{\cal H}^\pm_{\gamma \delta \gamma^{'} \delta^{'}}({\bf P},\Omega,{\bf q}) \cr
{\cal H}^{\pm \dagger}_{\gamma \delta \gamma^{'} \delta^{'}}
({\bf P},-\Omega,{\bf q}) &\delta_{\gamma \gamma^{'}}
\delta_{\delta  \delta^{'}}
{\cal G}^\pm({ \bf P},-\Omega,{\bf q}) } \right) \nonumber \\
\left( \matrix{ \langle {\bf q}|T^{\pm}_{\gamma^{'} \delta^{'} \alpha \beta}({\bf P},
\Omega)|{\bf p^{'}}\rangle  & \langle {\bf q}|L^{\pm}_{\gamma^{'} \delta^{'}\alpha \beta}
({\bf P},
\Omega)|{\bf p^{'}}\rangle  \cr
 \langle {\bf q}|L_{\gamma^{'} \delta^{'} \alpha \beta}^{\pm \ \dagger}({\bf P},
-\Omega)|{\bf p^{'}}\rangle  &
 \langle {\bf q}|T^{\pm \dagger}_{\gamma^{'} \delta^{'} \alpha \beta}({\bf P},
-\Omega)|{\bf p^{'}}\rangle  
 } \right)
\end{eqnarray}
where
\begin{eqnarray}
{\cal H}^{\pm}_{\gamma \delta \alpha \beta}({\bf P},
\Omega,{\bf p}) =
\int \frac{d\omega^{'}}{2 \pi}
\int \frac{d\omega^{''}}{2 \pi} \nonumber \\
\Big( F^<_{\alpha \gamma}({\bf P}/2+{\bf p},\omega^{''}-\omega^{'})
F^<_{\beta \delta}({\bf P}/2-{\bf p},
\omega^{'}) \nonumber \\ -F^>_{\alpha \gamma}({\bf P}/2+{\bf p},\omega^{''}-
\omega^{'})F^>_{\beta \delta}({\bf P}/2-{\bf p},\omega^{'}) \Big)/ 
\Big(\Omega -\omega^{''} \pm 
i\epsilon\Big)  \ 
\end{eqnarray}
denotes two disconnected anomalous propagators.
$L^\pm$ is the off-diagonal part of the generalized retarded $T$-matrix
(Fig. \ref{tmsffig}).

The anomalous part of the ladder ${\cal H}_{\alpha\beta\gamma\delta}$ 
does not mix
different values of the total spin (isospin) of the pair. 
It can be seen by writing the matrix structure of $H_{\alpha\beta}$ in
the basis of the total spin (isospin) of the pair and of its third component.
\begin{equation}
\label{h1}
{\cal H}=\left( \matrix{ (d_x+d_y)^2 & -i\sqrt{2} d_z(d_x+id_y)& -d_z^2 & 0 \cr
    -i\sqrt{2}d_x(d_x+id_y)& d_x^2+d_y^2-d_z^2& -i\sqrt{2}d_z(d_x-id_y)& 0\cr
-d_z^2 & -id_z(d_x-id_y) & (d_x-id_y)^2& 0 \cr
0 & 0 & 0 & D^2} \right) H
\end{equation}
for the triplet pairing with the matrix in the components $1^+, 1^0, 1^- ,
0^0$ of the total spin (isospin) of the pair and
\begin{equation}
\label{h2}
{\cal H}=\left( \matrix{ 0 & 0& -1 & 0 \cr
    0& 1& 0& 0\cr
-1 & 0 & 0& 0 \cr
0 & 0 & 0 &1} \right) H
\end{equation}
for the singlet pairing.
Using the structure of the anomalous Green's function (Eq. \ref{find})
and assuming a dependence of the angle averaged spin-isospin structure 
(the matrices in Eqs. \ref{h1}, \ref{h2})
  only on the value of  total  momentum
 one finds for central forces
\begin{equation}
\label{hsquare}
\sum_{\gamma \delta} {\cal H}_{\alpha^{'}\beta^{'}\gamma \delta} \times
{\cal H}_{\gamma \delta \alpha \beta} =
\delta_{\alpha^{'} \beta}\delta_{ \alpha \beta^{'}}{  H}^\dagger \times {  H} \ ,
\end{equation}
where $\times$ stands for momentum integrals with the 
interaction potential or the $T-$matrix
  as  occurring in the iteration of the Eq.
 (\ref{eqtl}).
Thus 
partial wave decomposition can be applied after angular averaging of the
intermediate uncorrelated two normal or two anomalous propagators in
 the ladder.
The resulting generalized
 $T$-matrix equation has a matrix structure corresponding to 
(Eq. \ref{eqtl}) in each partial wave 
\begin{eqnarray}
\label{eqtl2}
\left( \matrix{ \langle { p}|T^{\pm}_{(JST)}({ P},
\Omega)|{ p^{'}}\rangle  & \langle { p}|L^{\pm}_{(JST)}
({ P},
\Omega)|{ p^{'}}\rangle  \cr
 \langle { p}|L^{\pm}_{(JST)}({ P},
\Omega)|{ p^{'}}\rangle  &
 \langle { p}|T^{\pm \dagger}_{(JST)}({ P},
-\Omega)|{ p^{'}}\rangle  } \right) = \nonumber \\
\left( \matrix{ V_{(JST)}({ p},{ p^{'}})
& 0 \cr  0& V_{(JST)}({ p},{ p^{'}})
 } \right) 
+ \int \frac{q^2dq}{2 \pi^2}
\left( \matrix{ V_{(JST)}({ p},{ q})
& 0 \cr 0 &  V_{(JST)}({ p},{ q})
 } \right)\nonumber \\
 \left( \matrix{ 
{\cal G}^\pm({ P},\Omega,q) &
{\cal H}^\pm({ P},\Omega,q) \cr
{\cal H}^{\pm \dagger}
({ P},-\Omega,q) &
{\cal G}^\pm({ P},-\Omega,q) } \right) \nonumber \\
\left( \matrix{ \langle { q}|T^{\pm}_{(JST)}({ P},
\Omega)|{p^{'}}\rangle  & \langle { q}|L^{\pm}_{(JST)}
({ P},
\Omega)|{ p^{'}}\rangle  \cr
 \langle { q}|L^{\pm}_{(JST)}({P},
\Omega)|{ p^{'}}\rangle  &
 \langle { q}|T^{\pm \dagger}_{(JST)}({ P},
-\Omega)|{ p^{'}}\rangle  
 } \right) \ .
\end{eqnarray}
We used also $\langle p|L^\pm(P,\Omega)|q\rangle =\langle p|L^{\pm \ \dagger}(P,-\Omega)|q\rangle $.
In the following we skip the spin-isospin indices in the equations assuming a
partial wave decomposition in the generalized $T$-matrix.

\subsection{Self-energy}

The self-energy can be defined  generalizing the self-energy in the
$T$-matrix
approximation for the normal phase.
The diagonal part of the self-energy is defined by the 
$T$-matrix approximation to the two-particle Green's function (Fig. 
\ref{sigfig})
\begin{equation}
\label{sigmadt}
{\Sigma}^{<>}_{TM}(p,\omega)=i\int\frac{d\omega^{'}}
{2\pi}\frac{d^3k}{(2\pi)^3}
\langle {\bf(p-k)}/2| T_A^{<>}(\omega+\omega^{'},{\bf p+k})|{\bf(p-k)}/2\rangle 
 G^{<>}(k,\omega^{'}) \ \ .
\end{equation}
Analogously we can define the $T$-matrix part of the off-diagonal self-energy
(Fig. \ref{delfig})
\begin{equation}
\label{dtm}
{\Delta}^{<>}_{TM}(p,\omega)=i\int\frac{d\omega^{'}}
{2\pi}\frac{d^3k}{(2\pi)^3}
\langle {\bf(p-k)}/2| L_A^{<>}(\omega+\omega^{'},{\bf p+k})|{\bf(p-k)}/2\rangle 
 F^{<>}(k,\omega^{'}) \ \ .
\end{equation}
The above definition of self-energy is a $\Phi$-derivable approximation
\cite{lw} (Fig. \ref{phisf}).
It leads to thermodynamically consistent results. 
In Eqs. (\ref{dtm}) and (\ref{sigmadt}) the subscript $_A$ denotes the
antisymmetrization and we have not written explicitely the spin-isospin 
indices.
It can be checked that the formula (\ref{dtm}) conserves the singlet (triplet)
structure of the order parameter given by $F$.
The approximation (\ref{dtm}) does not include the usual BCS part of the
anomalous self-energy
\begin{equation}
\label{dbcs}
\Delta_{BCS}(p)=-i\int\frac{d\omega}{2 \pi}\int\frac{d^3 k}{(2 \pi)^3}
 V(p,k) F^{<}(k,\omega) \ \ .
\end{equation}
 This term can be added explicitely  to the the off-diagonal part of
the self-energy (Fig. \ref{delfig}).
 Moreover the $\Phi$ derivability of the approximation is
conserved.
As it will turn out the BCS part of the anomalous self-energy is dominant,
and in the first approximation one can neglect the two-body contribution
 (\ref{dtm})
 to the superfluid gap. However the approximation scheme consisting
 in keeping the $T$-matrix form of the diagonal self-energy and the BCS form
of the off-diagonal self-energy is not $\Phi$-derivable.

\subsection{Quasi-particle  approximation}

The full solution of the set of equations (\ref{eqtl}) for the generalized
$T$-matrix, the normal self-energy and the superfluid gap
requires a self-consistent
 iterative solution.
Below we present results of a simpler calculation. It starts with the mean
field approximation for the normal self-energy (Hartree-Fock approximation)
and the BCS approximation for the superfluid gap
\begin{equation}
\label{bcs}
 \Delta_{BCS}(p) = 
- \int \frac{d^3k}{(2\pi)^3}V(p,k)\frac{
\left(1-2f(E_k)\right)}{2E_k}\Delta_{BCS}(k) \ .
\end{equation}
The starting  spectral function for the calculation of the $T$-matrix is
\begin{equation}
\label{spec1}
A_{BCS}(p,\omega)=2 \pi \left(\frac{E_p+\xi_p}{2E_p}\delta(\omega-E_p)+
\frac{E_p-\xi_p}{2E_p}\delta(\omega+E_p)\right)
\end{equation}
where $\xi_p=p^2/2m+\Sigma_{HF}(p)-\mu$ and $E_p=\sqrt{\xi_p^2+\Delta_p^2}$.
The corresponding off-diagonal spectral function is
\begin{equation}
\label{spec2}
B(p,\omega)=-2\pi \frac{\Delta_p}{2E_p} \left( \delta(\omega-E_p)-
\delta(\omega+E_p)\right) \ .
\end{equation}

The generalized $T$-matrix equation is solved for the $S$-wave Yamaguchi 
interaction \cite{yamaguchi}.
 This oversimplified interaction is used to reduce the numerical
difficulties involved in the solution of $T$-matrix equations and in the
calculation of the self-energies.
First the imaginary parts of the diagonal  and off-diagonal self-energies 
are calculated
\begin{eqnarray}
{\rm Im}\Sigma(p,\omega)=\int \frac{d^3k}{(2\pi)^3}\left(\frac{E_p+\xi_p}
{2E_p}\left(f(E_p)+b(\omega+E_p)\right) \right. \nonumber \\
\langle ({\bf p-k})/2|{\rm Im}T(
{\bf p+k},\omega+E_p)| ({\bf p-k})/2\rangle _A \nonumber \\
\left. +\frac{E_p-\xi_p}
{2E_p}\left(f(-E_p)+b(\omega-E_p)\right)\langle ({\bf p-k})/2|{\rm Im}T(
{\bf p+k},\omega-E_p)| ({\bf p-k})/2\rangle _A \right) 
\end{eqnarray}
and similarly for ${\rm Im}\Delta_{TM}$ ($b(\omega)$ is the Bose
distribution).
 Then from the dispersion relations the real part of the
dispersive contribution to the self-energy can be obtained
\begin{equation}
\label{dispd}
\left(\matrix{ {\rm Re}\Sigma(p,\omega)_{disp} 
\cr {\rm Re} \Delta(p,\omega)_{TM}}\right)
={\cal P}\int\frac{d\omega^{'}}{2\pi}\left(
\matrix{{\rm Im}\Sigma(p,\omega^{'})_{disp} \cr {\rm Im}
\Delta(p,\omega^{'})_{TM}}\right)
\frac{1}{\omega-\omega^{'}} \ .
\end{equation}
 To the dispersive
part of the self-energies one has to add the mean-field self-energy
$\Sigma_{HF}$ and the mean
field superfluid gap $\Delta_{BCS}$
\begin{eqnarray}
{\rm Re}\Sigma(p,\omega)= {\rm Re}\Sigma(p,\omega)_{disp}+\Sigma_{HF}(p)
\nonumber \\
{\rm Re}\Delta(p,\omega)= {\rm Re}\Delta(p,\omega)_{TM}+\Delta_{BCS}(p)
\end{eqnarray}
that have been calculated already  to obtain the
spectral functions  (\ref{spec1}), (\ref{spec2}).

\section{Results for the $T$-matrix}

\subsection{Energy gap in the $T$-matrix and the two-particle 
spectral function}

The equations for the generalized $T$-matrix are solved with the
quasi-particle ansatz for the spectral functions $A$ and $B$.
We present results at the temperature of $3$MeV with the BCS gap of 
$10.1$MeV.
The results do not
change appreciably when reducing the temperature further. 
In Fig \ref{imtfig} is shown the imaginary part of the diagonal and
off-diagonal parts of the generalized $T$-matrix. From the BCS ansatz for the
spectral functions follows a  gap in two-particle excitations around the 
the Fermi energy. For zero total momentum of the pair, this forbidden region
 is twice the superfluid gap. At the edge of the two-particle gap the
 generalized $T$-matrix has a singularity as function of energy, similar to the
 singularities in the one-particle spectral function in the BCS approximation.

In fact the imaginary part of the  $T$-matrix is related to the two-particle
propagator in the $T$-matrix approximation.
The one-time two-particle spectral function is
\begin{eqnarray}
\left( \matrix{ \langle { p}| A_2({ P},
\Omega)|{ p^{'}}\rangle  & \langle { p}|B_2
({ P},
\Omega)|{ p^{'}}\rangle  \cr
 \langle { p}|B_2({ P},
-\Omega)|{ p^{'}}\rangle  &
 \langle { p}|A_2({ P},
-\Omega)|{ p^{'}}\rangle  } \right) = \nonumber \\ -2 {\rm Im}
\int \frac{d^3 q}{(2\pi)^3}
\int \frac{d^3 k}{(2\pi)^3}
\left( \matrix{ V^{-1}({ p},{ p^{'}})
& 0 \cr  0& V^{-1}({ p},{ p^{'}})
 } \right) \nonumber \\
\left( \matrix{ \langle { q}|T^{\pm}({ P},
\Omega)|{k}\rangle  & \langle { q}|L^{\pm}
({ P},
\Omega)|{ k}\rangle  \cr
 \langle { q}|L^{\pm}({ P},
-\Omega)| k\rangle  &
 \langle { q}|T^{\pm \dagger}({P},
-\Omega)| k\rangle  
 } \right)
\left( \matrix{ V^{-1}({ k},{ p^{'}})
& 0 \cr 0 &  V^{-1}({ k},{ p^{'}})
 } \right)
\end{eqnarray}
The spin-isospin indices are omitted, and the two-particle spectral function
can be projected on a definite total spin or isospin of the pair.
Also the relative momentum can be projected on states with 
definite angular momentum. In Fig. \ref{g2fig} we present 
the two particle spectral function projected on partial waves occurring in 
our
separable interaction 
\begin{equation}
\langle A_2(P,\Omega)\rangle  = \int\frac{ d^3k d^3 p}{(2\pi)^6}
 g(k)g(p) \langle k|A_2(P,\Omega)|p\rangle  \ ,
\end{equation}
where $g(k)=1/(k^2+\beta^2)$
 is the formfactor of the Yamaguchi
interaction \cite{yamaguchi}.
We use the same projection for the off-diagonal two-particle spectral function
$B_2$ in different channels and for the 
uncorrelated BCS two-particle propagators.

 The energy gap present in the two-particle BCS propagator is visible also in
 the $T$-matrix approximation (Fig. \ref{g2fig}). 
It corresponds to a minimal energy
 to excite a two-particle pair of twice the single-particle energy gap.
A similar gap appears in the two-particle anomalous propagator (Fig.
 \ref{b2fig})
and for nonzero momentum of the pair (Figs. \ref{gg2fig}, \ref{bb2fig}).

\subsection{Singularity in the $T$-matrix}
\label{tsing}

The $T$-matrix is singular for the total energy of the pair equal to twice
the Fermi energy and zero total momentum. It is a generalization of the
Thouless criterion for the critical temperature to the superfluid phase.
Indeed the imaginary part of $T$ and $L$ is always zero at $\Omega=0$ and
$P=0$ and the inverse of the real part of the generalized $T$-matrix
\begin{equation}{\rm Re}
\langle k^{'}|\left( \matrix{ T & L \cr
L & T^\dagger} \right)_{P=0,\Omega=0}|k\rangle ^{-1}
\end{equation}
has a zero eigenvalue
\begin{equation}\int d^3k
\langle k^{'}|\left( \matrix{ T & L \cr
L & T^\dagger }\right)_{P=0,\Omega=0}|k\rangle ^{-1}
\left(\matrix{ \Delta(k) \cr -\Delta(k) }\right)=0 \ .
\end{equation}
$\Delta(k)$ (with spin-isospin indices omitted) is a 
solution of the mean-field gap equation
\begin{equation}
 \Delta(p) - 
 \int \frac{d^3k}{(2\pi)^3}V(p,k)\frac{d\omega d\omega^{'}}{(2\pi)^2}
\frac{A(\omega,k)A_s(\omega,k)}{\omega+\omega^{'}}
\left(1-f(\omega)-f(\omega^{'}\right))\Delta(k)=0
\end{equation}
if $A_s$ is the full spectral function with only the BCS contribution to 
the off-diagonal self-energy $\Delta_{BCS}$.
In particular the inverse generalized $T$-matrix with
 mean-field quasi-particle propagators (\ref{spec1}), (\ref{spec2}) has 
a zero eigenvalue corresponding to the BCS mean-field superfluid gap 
(\ref{bcs}). In Fig. \ref{retfig} is shown the real part of the inverse
determinant of the generalized $T$-matrix for the two $S$ partial waves.
Clearly the superfluid gap is formed in the $^3S_1$ channel 
(for the chosen interaction). The generalized $T$-matrix in this channel
shows a singularity at zero total momentum and at twice the Fermi energy
for all temperatures below $T_c$.

\section{Self-energies and spectral functions}

\subsection{Single-particle energies} 

The position of the single particle pole $\omega_p$ 
when including only the real part of
the self-energy can be obtained from the solution of 
\begin{equation}
\label{speq}
\omega_p=\frac{p^2}{2 m} +{\rm Re}\Sigma(p,\omega_p) -\mu \ .
\end{equation}
In Fig. \ref{omfig} it is compared to the position of the quasi-particle 
pole $\xi_p$ when including
only the Hartree-Fock self-energy.
There is a significant difference between the  mean-field single-particle
energy and the 
 single-particle energy (\ref{speq}) for low-momenta, 
due to the dispersive part of the self-energy as obtained in
the $T$-matrix approximation. 
 It is a reflection of the presence of
short range correlations for nuclear interactions. This modification of 
single-particle energies below the Fermi energy leads to important corrections 
to the binding energy per particle in nuclear matter.
A resummation of ladder diagrams in the form of the $G$-matrix (or the
$T$-matrix) is necessary to obtain reliable results for the binding energy
\cite{day,walecka}.

In the superfluid the presence of the off-diagonal self-energy
$\Delta(p,\omega)$ leads to the splitting of quasi-particle poles in the 
spectral function $A_s(p,\omega)$
\begin{equation}
\label{epeq}
E_p=\pm\sqrt{(\xi_p+{\rm Re}\Sigma(p,E_p))^2+|\Delta(p,E_p)|^2} \ .
\end{equation}
As shown in Fig. \ref{omfig} the superfluid quasi-particles, present a gap in
excitations around the Fermi momentum. Far from from the Fermi momentum the
dominant quasi-particle pole of $A_s(p,\omega) $ approaches
the quasi-particle pole $\omega_p$ of $A(p,\omega)$.
In Fig. \ref{eomfig} the position of the quasi-particle poles obtained
including the full off-diagonal self-energy (Eq. \ref{epeq}) is compared to one
where only the BCS part of the superfluid gap is taken   
\begin{equation}
\label{epeqbcs}
E_p=\pm\sqrt{(\xi_p+{\rm Re}\Sigma(p,E_p))^2+|\Delta_{BCS}(p)|^2} \ .
\end{equation}
The two energies are very close. 
At the scale of Fig. \ref{omfig} they cannot
be distinguished and only close to the Fermi momentum a small difference 
can be seen in Fig. \ref{eomfig}.
It is due to the fact that the dispersive part of the off-diagonal
self-energy $\Delta_{TM}(p,\omega)$ introduces only a small correction to the
mean-field superfluid gap $\Delta_{BCS}(p)$.

\subsection{Imaginary part of the self-energy}

In Fig. \ref{gamfig} we plot the energy dependence of 
$-{\rm Im}\Sigma^{+}(p,\omega)$ for several values of the momentum $p$. 
A characteristic feature is the very strong reduction of the damping in 
the region around the Fermi energy, already at $ T=3$MeV. 
It is related to the appearance of an energy gap for excitations of twice the
value of the superfluid gap. On top of that the usual reduction of 
 damping due to the
restriction of the phase space appears, leading to the formation of
quasi-particles around the Fermi surface.

The quasi-particle nature of the excitations can be judged by plotting the 
        single particle  width
\begin{equation}
\Gamma(\omega,p)=-2{\rm Im}\Sigma(p,\omega) \ 
\end{equation}
at the quasi-particle pole.
The single particle width is small around the Fermi energy for the 
quasi-particle pole $\omega=\omega_p$ and the poles $\omega=\pm E_p$ 
(Fig. \ref{widthfig}).
The formation of the energy gap for excitation of
particle pairs and the reduction of the phase space around the Fermi surface
leads to the appearance of sharp quasi-particles for momenta close to the 
Fermi momentum.

\subsection{Superfluid gap}

The superfluid gap acquires a contribution  $\Delta_{TM}$
 from the $T$-matrix diagram.
 The real part of the $T$-matrix gap is  obtained from the
 dispersion relation (\ref{dispd}). In Fig. \ref{deltrfig} is shown ${\rm Re}
\Delta_{TM}(p,\omega)$ for several values of momenta. Its value is small 
for energies corresponding to the Fermi energy and for momenta close 
to the Fermi
 momentum.

In Fig. \ref{deltifig}{ is shown the imaginary part of the $T$-matrix
  contribution to the superfluid gap.
The imaginary part of the dispersive contribution to the superfluid gap 
 ${\rm Im}\Delta_{TM}(p,\omega)$ shows a gap around the Fermi energy similarly
  as the imaginary part of the diagonal self-energy.
This means that the imaginary part is not modifying the value of the
  superfluid gap around the Fermi surface.

In Fig. \ref{deltfig} the mean-field superfluid gap $\Delta_{BCS}(p)$
is compared to the superfluid gap containing also the $T$-matrix contribution
at the quasi-particle poles $\omega_p$ or $E_p$. Close to the Fermi momentum
the mean-field value of the superfluid gap is not modified significantly.
Only for small momenta the $T$-matrix contribution to $\Delta(p,E_p)$ shows up.

\subsection{Spectral functions}

The quasi-particle nature of the excitations around the Fermi momentum can be
judged from the spectral functions $A(p,\omega)$ and $A_s(p,\omega)$.
In the upper panel of Fig. \ref{specfig} are shown the spectral functions for 
$p=0$ (below the Fermi surface). The main strength of the spectral function is
concentrated around the quasi-particle peak at 
$\omega=-56{\rm MeV}=\omega_p\simeq-E_p$. However for this momentum the
spectral function is relatively spread. The quasi-particle approximation
would be of limited validity. Above the Fermi energy the strength of the
spectral function is small. It is slightly larger for $A_s$ than for $A$
because of the
contribution of the second quasi-particle pole of the full spectral function
$A_s(p,\omega)$ in the superfluid. 

The situation is reversed in the lower panel of Fig. \ref{specfig} where the 
spectral functions are plotted for the momentum $p=350$MeV (above the Fermi 
momentum). The quasi-particle pole of $A$ and the dominant quasi-particle pole
of $A_s$ are located above the Fermi energy. Below the Fermi energy 
there is a small contribution
from the background strength of the spectral
function to $A$ and $A_s$ and a small contribution from the second pole to the
full spectral function $A_s$.

In the middle panel of Fig \ref{specfig} are shown the spectral functions for
a momentum close to the Fermi momentum.
Very sharp quasi-particle peaks are visible for $A(p,\omega)$ 
at $\omega=\omega_p$
and for $A_s(p,\omega)$ at $\omega=\pm E_p$. For this momentum the difference
between $A$ and $A_s$ is the most pronounced, because $\omega_p$ is different
from $E_p$ and both poles of $A_s$ have comparable strength.

In Fig. \ref{aspecfig} is plotted the off-diagonal spectral function 
$B(p,\omega)$. The anomalous spectral function is an odd function of energy. 
$B(p,\omega)$ has two quasi-particle poles on both sides of the
Fermi energy. The quasi-particle poles are very sharp for momenta close to the
Fermi momentum. Only for small momenta the off-diagonal spectral function
shows two relatively broad peaks.

\section{Conclusions}

We  present an approach which allows for the resummation of  ladder
diagrams in the form of the in-medium $T$-matrix and the treatment of the
superfluid phase of the nuclear matter at the same time. The approximation
presented in Sect. \ref{superladder} is a generalization of the usual
$T$-matrix resummation in medium to temperatures below $T_c$. In the superfluid
phase the Green's functions, self-energies and the $T$-matrix acquire
additional indices corresponding to anomalous propagators.
The approach is $\Phi$-derivable and hence it is thermodynamically consistent,
like the self-consistent $T$-matrix approximation. 
The partial wave expansion can be performed for central forces.
The corresponding self-energy is calculated assuming the $T$-matrix
approximation for the two-particle propagator and the two-particle  anomalous
propagator. The off-diagonal self-energy can be supplemented with the
mean-field BCS contribution which turns out to be 
 dominant. The addition of the BCS gap to
the off-diagonal self-energy
does not spoil the $\Phi$ derivability of the approximation.
The spin-isospin structure of the $T$-matrix part of the superfluid order
parameter is the same  as its mean-field part.

In this  exploratory work we use a simple separable interaction to illustrate 
the method by numerical results.
The calculations are performed in the approximation where the normal and
anomalous propagators in the generalized $T$-matrix ladder and in the
self-energy diagrams  are of the BCS mean-field form. It represents the first
iteration  in the calculation of the self-consistent set of equations with 
off-shell normal and anomalous propagators. It turns out that the real part
of
the diagonal
  self-energy changes in the first iteration significantly from its
mean-field form. On the other hand the value of the superfluid gap is only
slightly modified by the addition of the $T$-matrix contribution in the first 
iteration. 
 The imaginary part of the diagonal self-energy is reduced around the
Fermi energy due to the appearance of an energy gap for excitation of twice
the value of the superfluid gap. The quasi-particle approximation is a
reasonable description of excitations close to the Fermi surface, with 
single-particle energies modified by the resummation of ladder diagrams.

Let us comment on the influence of the use of
 more realistic forces on the results.
The importance of the ladder diagram resummation for the real-part of the
self-energy is of course important 
for any realistic interactions \cite{day,walecka,roepke,Schnell,ja}.
The value of the superfluid gap is expected to be much smaller for realistic
nuclear potentials at normal nuclear density. The validity of the partial wave
expansion for non-central forces in the superfluid remains to be checked.
However since we expect that the dominant part of the off-diagonal self-energy
comes from BCS diagram the partial wave expansion in the generalized
$T$-matrix calculation would not introduce important modifications for final
self-energies.

Another open question is the effect of the self-consistency of the calculation
on the spectral
functions and self-energies. The equations for the
self-energies and the generalized $T$-matrix should be iterated instead 
of using
only the quasi-particle mean-field propagators. In the self-consistent 
iteration 
the superfluid gap is reduced \cite{ja3} and the energy gap in the superfluid
is no longer sharp. The
 imaginary part of the self-energy will also be reduced
 around the Fermi energy but not as sharply as in Fig. \ref{widthfig}.
We note that only the use of self-consistent propagators and self-energies in
the generalized $T$-matrix approximation guarantees the thermodynamic
consistency of the results \cite{ja4}. 
As a result of the self-consistency in the off-diagonal self-energy the 
 generalized $T$-matrix  will not have a singularity at twice the Fermi
energy and zero total momentum. As described in Sect. \ref{tsing} this
singularity is present only if mean-field BCS off-diagonal self-energy
is used in propagators of the ladder. 
The inclusion of two-particle contribution to the order parameter
modifies this criterion for long range order in the two-particle propagator.

Finally comparing to results of Ref. \cite{ja2} we found a similar behavior
of the real part of the generalized $T$-matrix if using BCS propagators,
i.e. a singularity at twice the Fermi energy and zero total momentum. However
the imaginary part of the $T$-matrix is very different form the results in
Ref. \cite{ja2}. The generalized
$T$-matrix approximation
shows an excitation
gap around the Fermi energy, which leads to similar gaps in the one-particle
and two-particle spectral functions. This feature 
is very comforting since it is a
manifestation of the energy gap  for the excitation of
particle pairs present in the superfluid.

\vskip .3cm
This work was partly supported by the KBN
under Grant 2P03B02019.


\newpage

\begin{figure}

\centering
\includegraphics[width=0.9\textwidth]{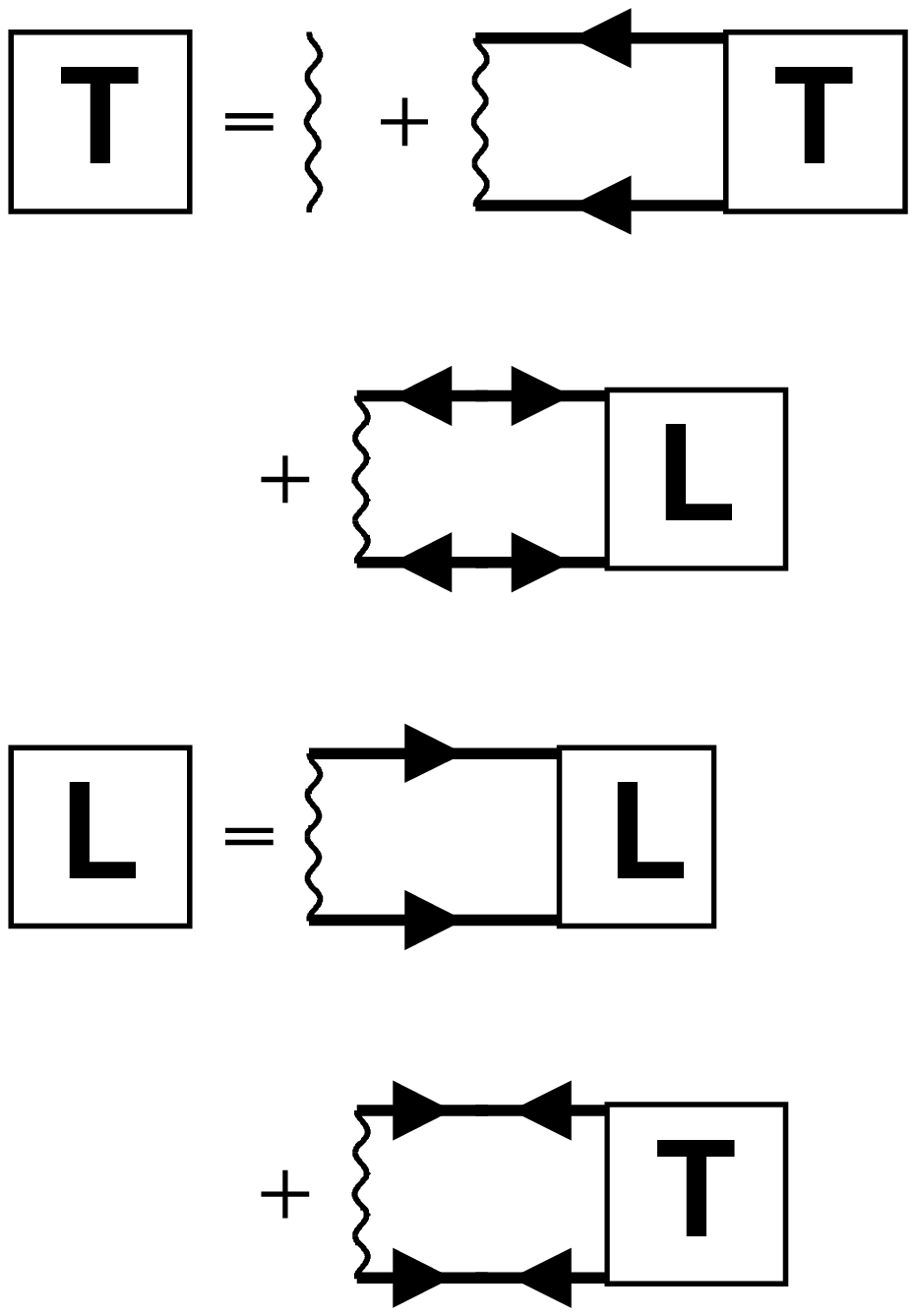}
\caption{The generalized $T$-matrix equation in the superfluid.}
\label{tmsffig}
\end{figure}

\begin{figure}

\centering
\includegraphics[width=0.9\textwidth]{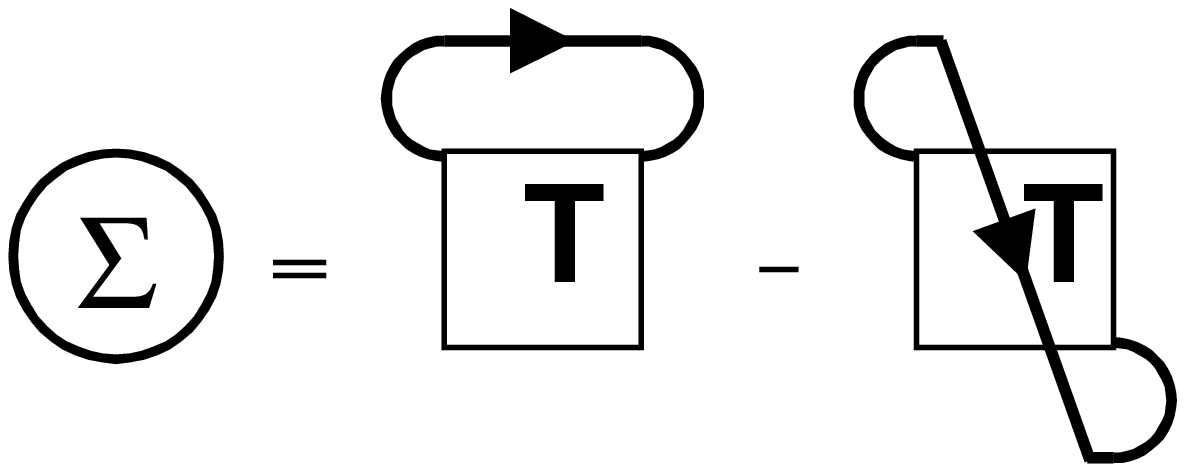}
\caption{The normal part of the self-energy
  in the $T-$matrix approximation. }
\label{sigfig}
\end{figure}

\begin{figure}

\centering
\includegraphics[width=0.9\textwidth]{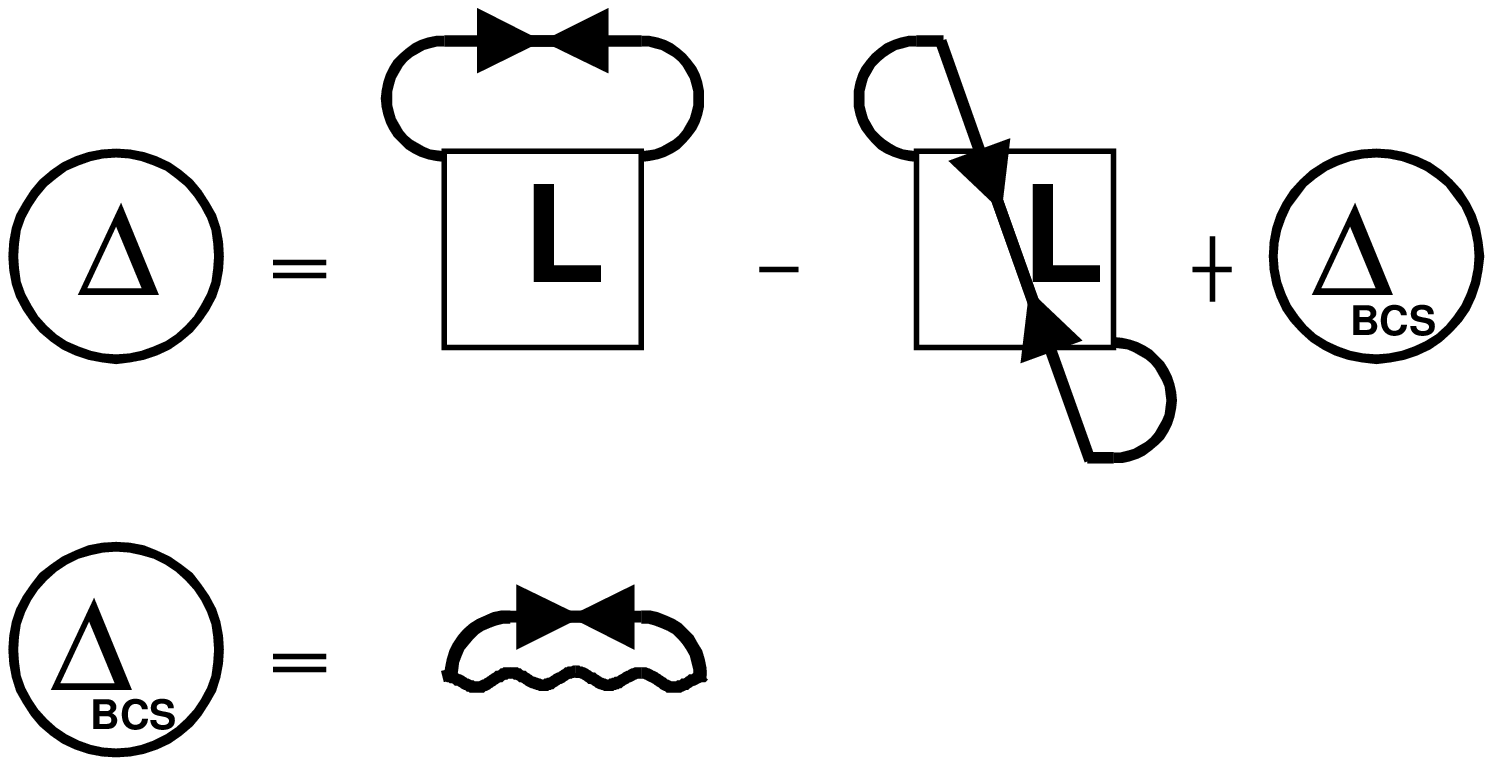}
\caption{The off-diagonal part of the self-energy
  in the generalized $T-$matrix approximation with the BCS contribution.}
\label{delfig}
\end{figure}

\begin{figure}

\centering
\includegraphics[width=0.9\textwidth]{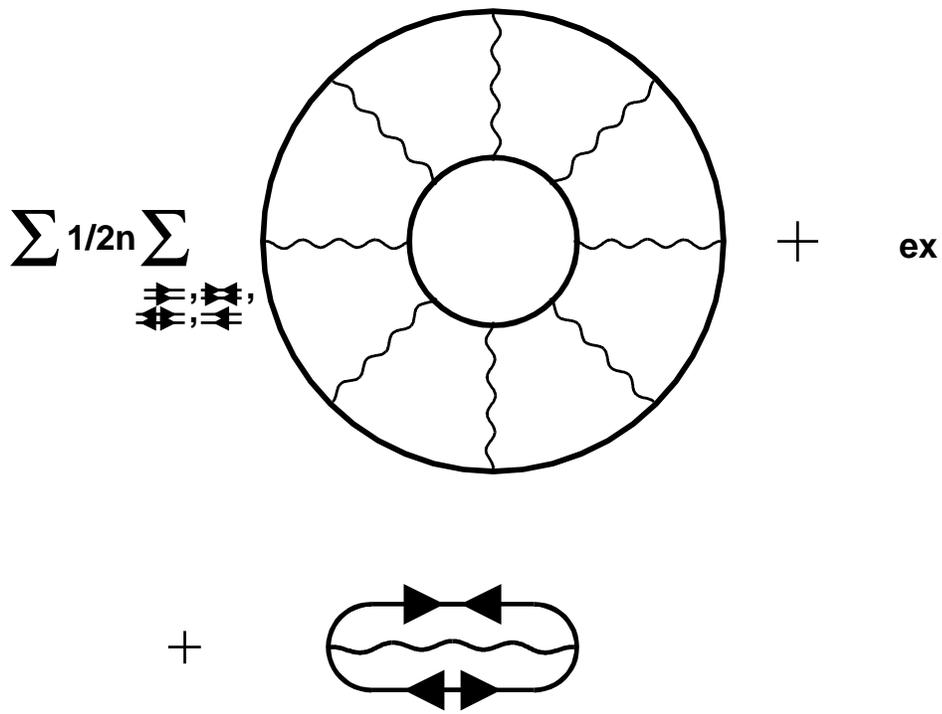}
\caption{Diagrams contributing to the generating functional $\Phi$ 
 in the generalized $T-$matrix approximation. }
\label{phisf}
\end{figure}

\begin{figure}

\centering
\includegraphics[width=0.9\textwidth]{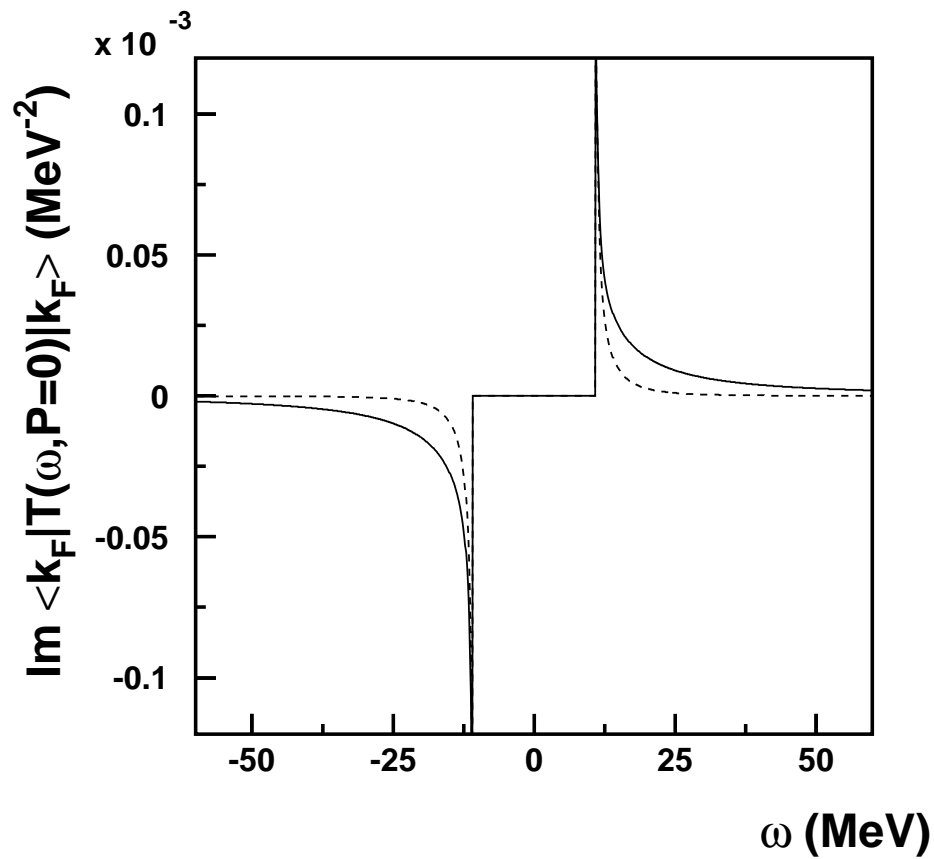}
\caption{Imaginary part of the diagonal (solid line) and off-diagonal
(dashed line) part of the generalized $T$-matrix             in the 
$^3S_1$ channel}
\label{imtfig}.
\end{figure}

\begin{figure}

\centering
\includegraphics[width=0.8\textwidth]{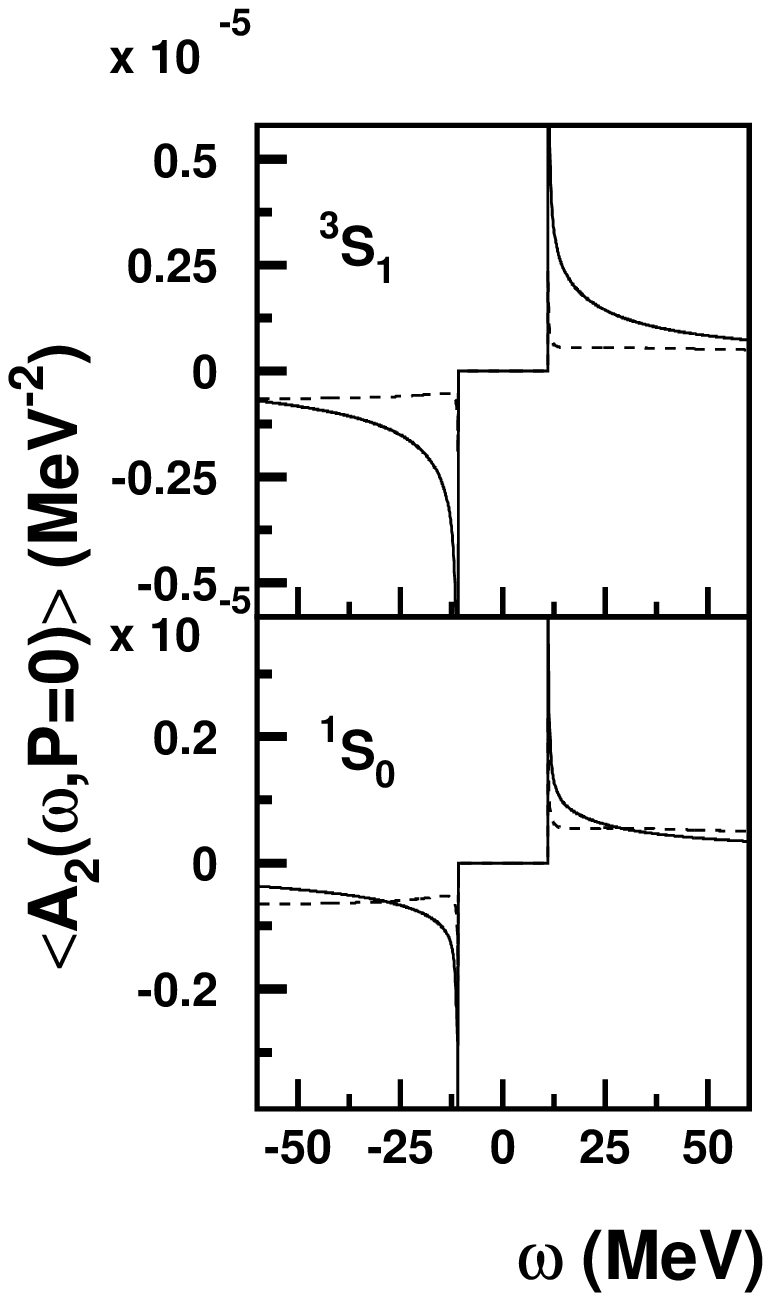}
\caption{Two-body spectral function for normal propagators obtained form the
  generalized $T$-matrix (solid line) and from two uncorrelated BCS propagators
(dashed line).
Projected on the $^3S_1$ channel (upper panel) and  $^1S_0$
 channel (lower panel).}
\label{g2fig}
\end{figure}

\begin{figure}

\centering
\includegraphics[width=0.8\textwidth]{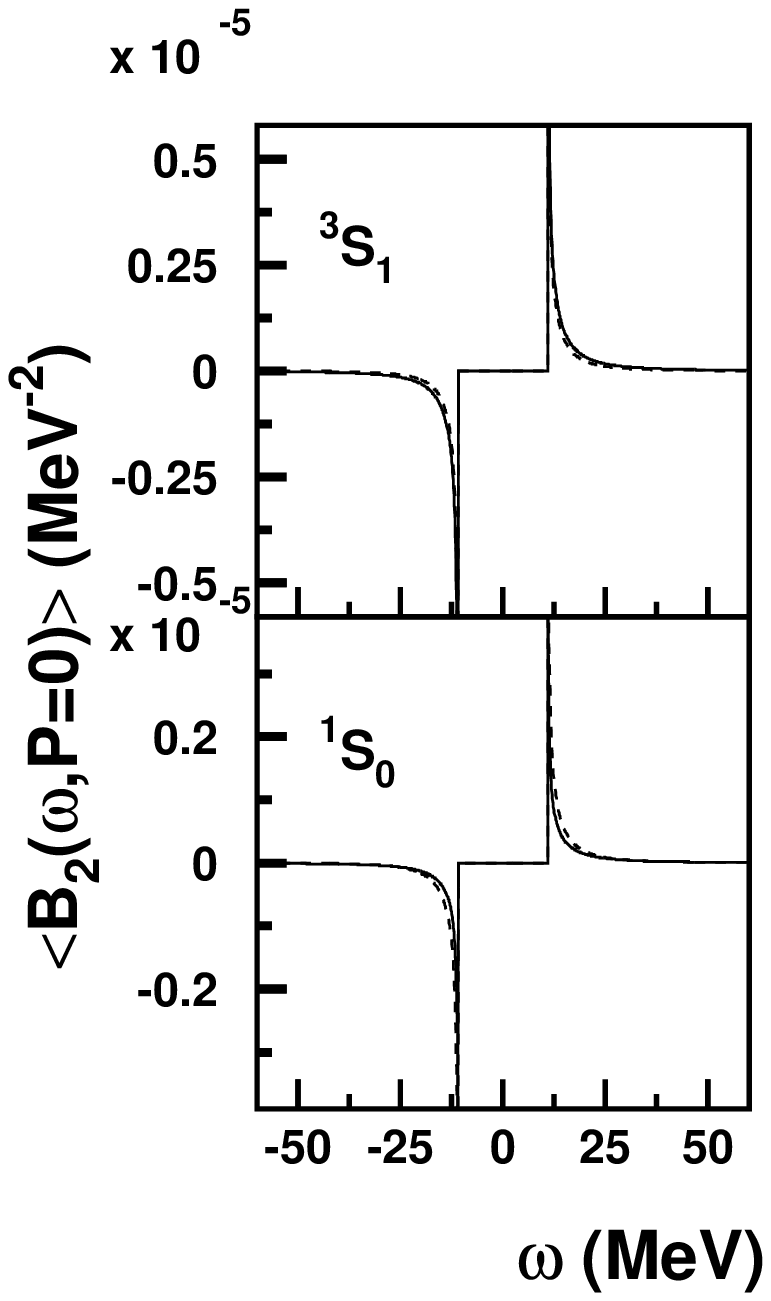}
\caption{Two-body spectral function for anomalous propagators 
obtained form the
  generalized $T$-matrix (solid line) and from two uncorrelated BCS anomalous
 propagators (dashed line).
Projected on the $^3S_1$channel (upper panel) and  $^1S_0$
channel (lower panel).}
\label{b2fig}
\end{figure}

\begin{figure}

\centering
\includegraphics[width=0.9\textwidth]{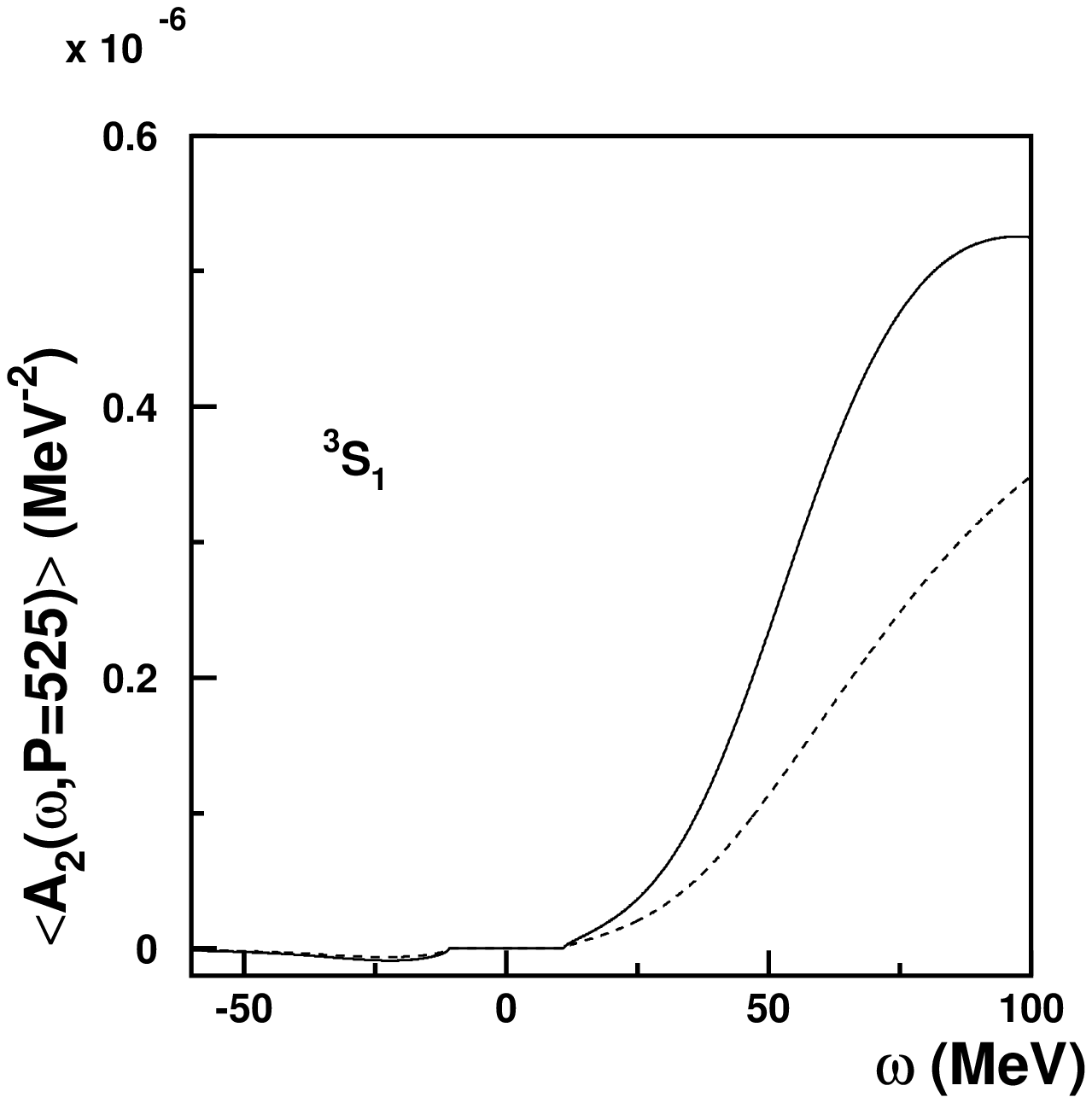}
\caption{Two-body spectral function for normal propagators
obtained form the
  generalized $T$-matrix (solid line) and from two uncorrelated BCS 
 propagators (dashed line) for nonzero total momentum of the pair ($P=525$MeV),
projected on the $^3S_1$ channel.}
\label{gg2fig}
\end{figure}
\begin{figure}

\centering
\includegraphics[width=0.9\textwidth]{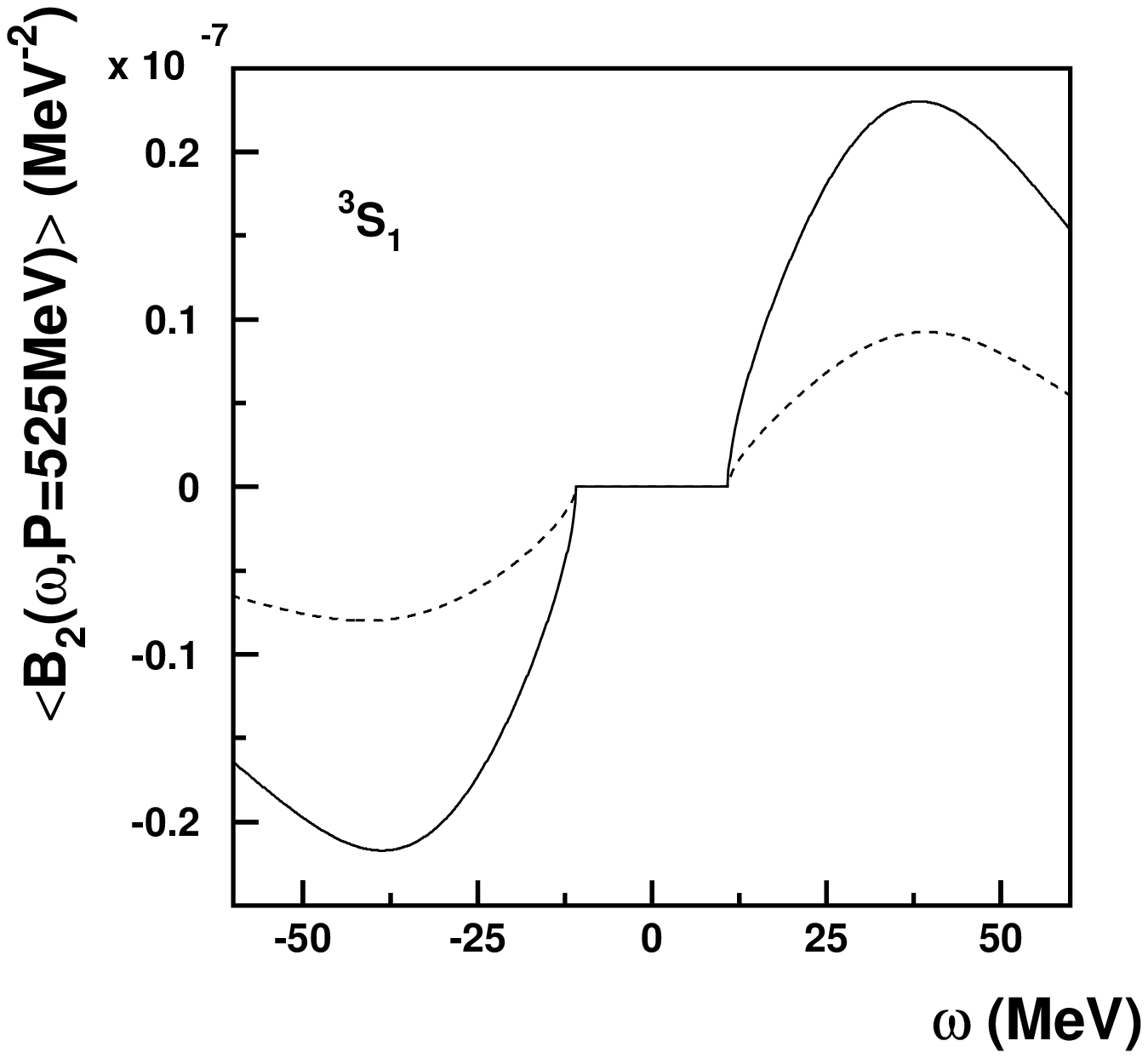}
\caption{Two-body spectral function for anomalous propagators 
obtained form the
  generalized $T$-matrix (solid line) and from two uncorrelated BCS 
 anomalous propagators (dashed line)
 for nonzero total momentum of the pair ($P=525$MeV),
projected on the $^3S_1$ channel.}
\label{bb2fig}
\end{figure}

\begin{figure}

\centering
\includegraphics[width=0.9\textwidth]{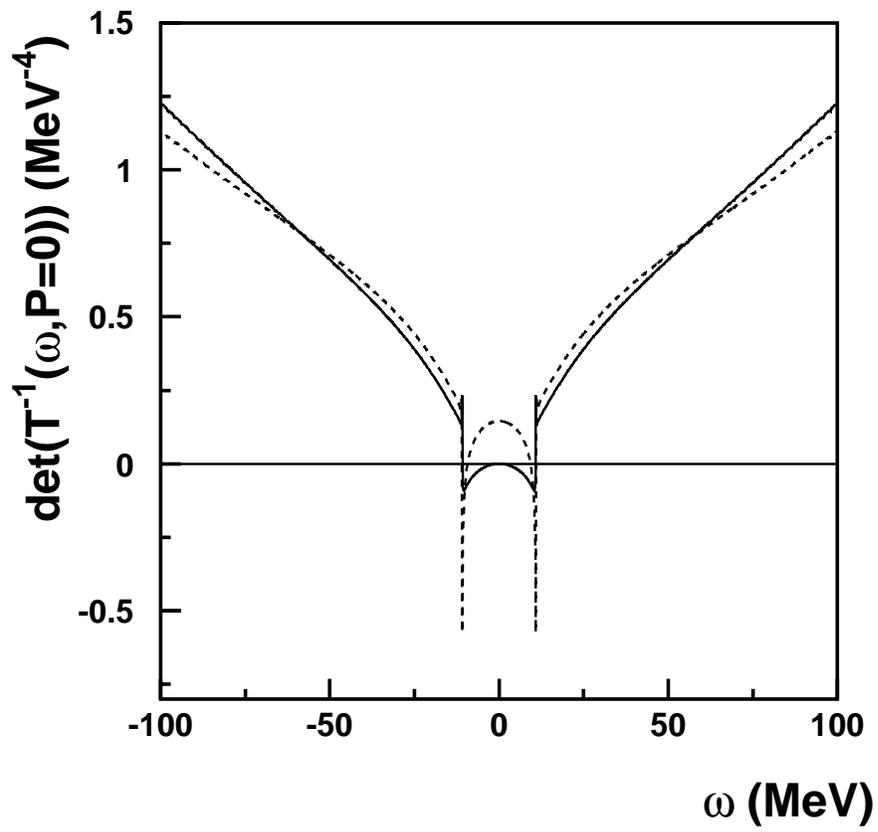}
\caption{Real part of the determinant of the inverse $T$-matrix
for the $^3S_1$ (solid line) and the $^1S_0$
 channel (dashed line).}
\label{retfig}
\end{figure}

\begin{figure}

\centering
\includegraphics[width=0.9\textwidth]{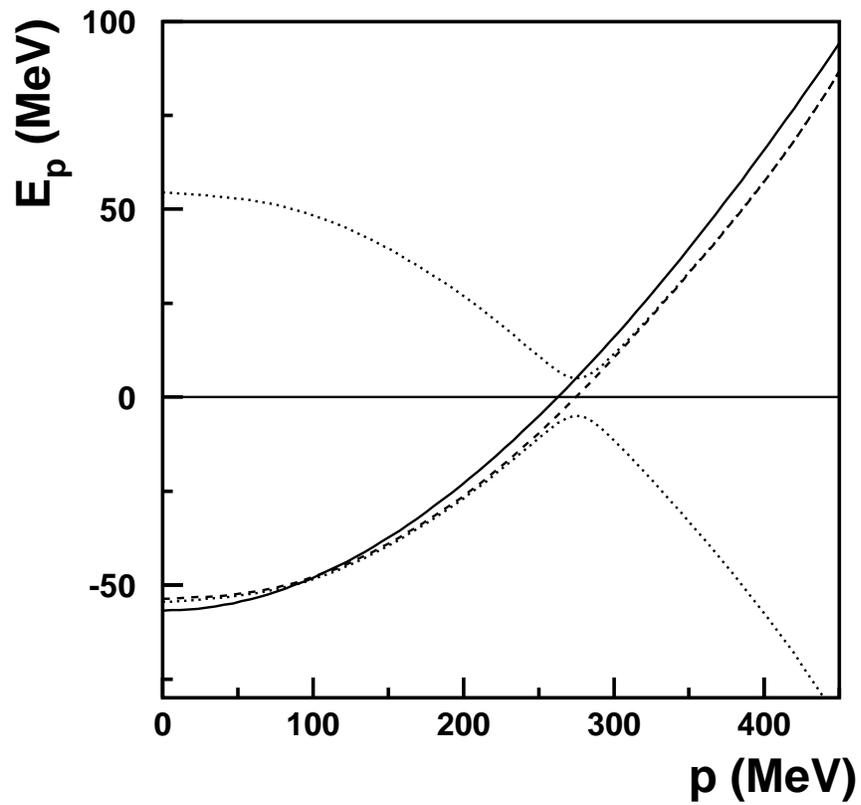}
\caption{The position of the quasi-particle pole for normal self-energy
for the HF propagator (solid line) and for the propagator including the normal
self-energy from the $T$-matrix calculation (dashed line). 
The positions of quasi-particle
  poles of the full propagator including the normal and the off-diagonal
  self-energies   in the $T$-matrix approximation are denoted by the dotted
 lines.}
\label{omfig}
\end{figure}

\begin{figure}

\centering
\includegraphics[width=0.9\textwidth]{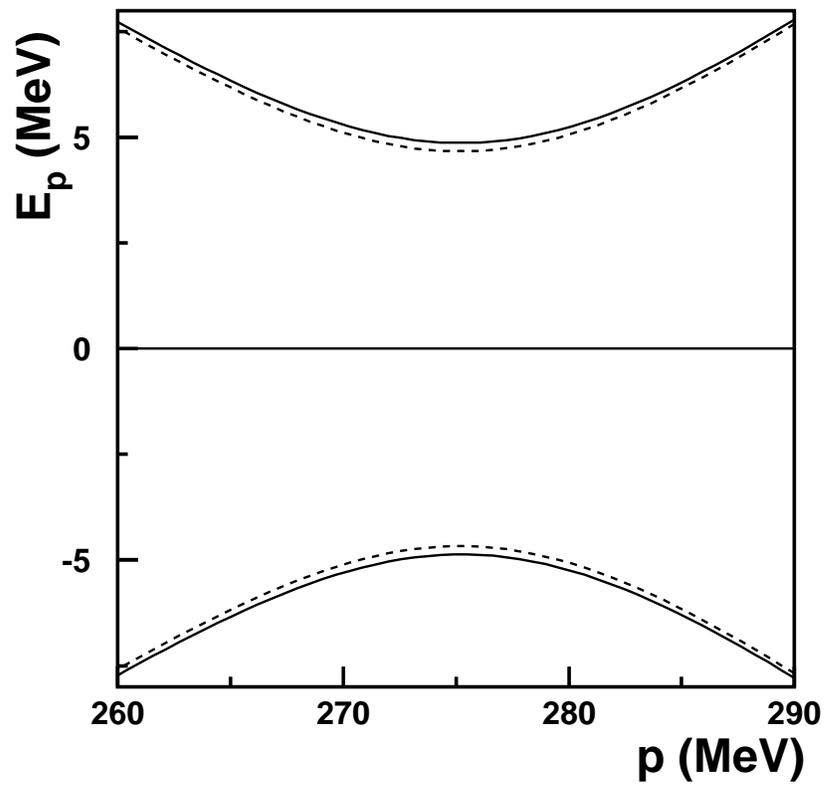}
\caption{The positions of quasi-particle
  poles of the full propagator including the normal and the off-diagonal
  self-energies   in the $T$-matrix approximation are denoted by the dashed
 lines. The solid line represents the positions of the poles when taking the
  BCS form only for the off-diagonal self-energy.}
\label{eomfig}
\end{figure}

\begin{figure}

\centering
\includegraphics[width=0.9\textwidth]{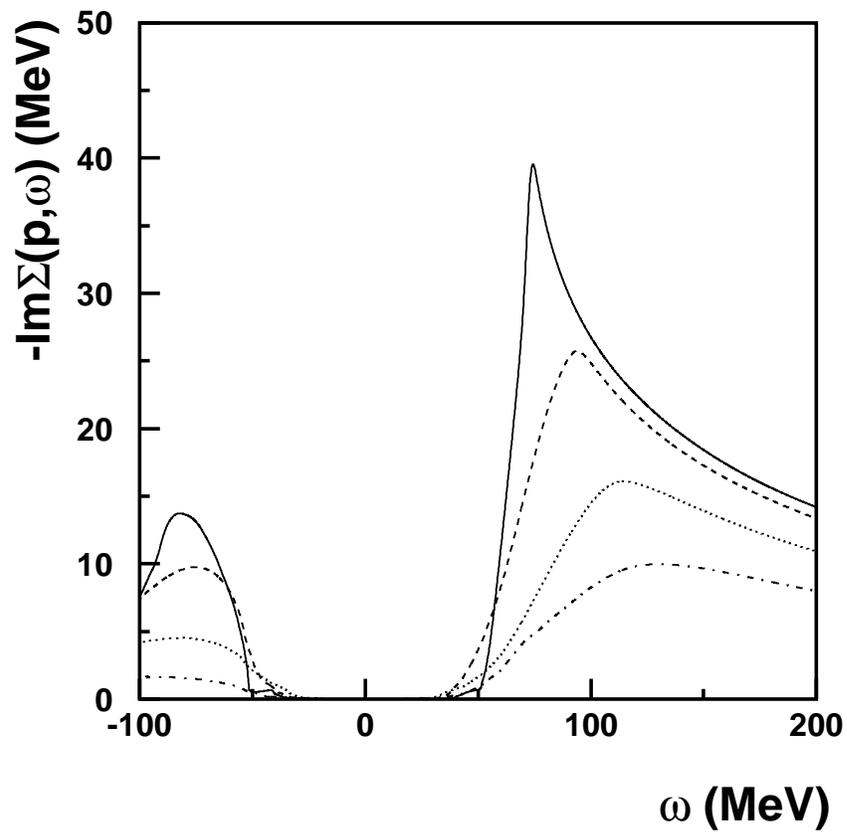}
\caption{The 
imaginary part of the retarded self-energy as function of energy for
$p=0,140,280,420$MeV (solid, dashed, dotted, dash-dotted lines). }
\label{gamfig}
\end{figure}

\begin{figure}

\centering
\includegraphics[width=0.9\textwidth]{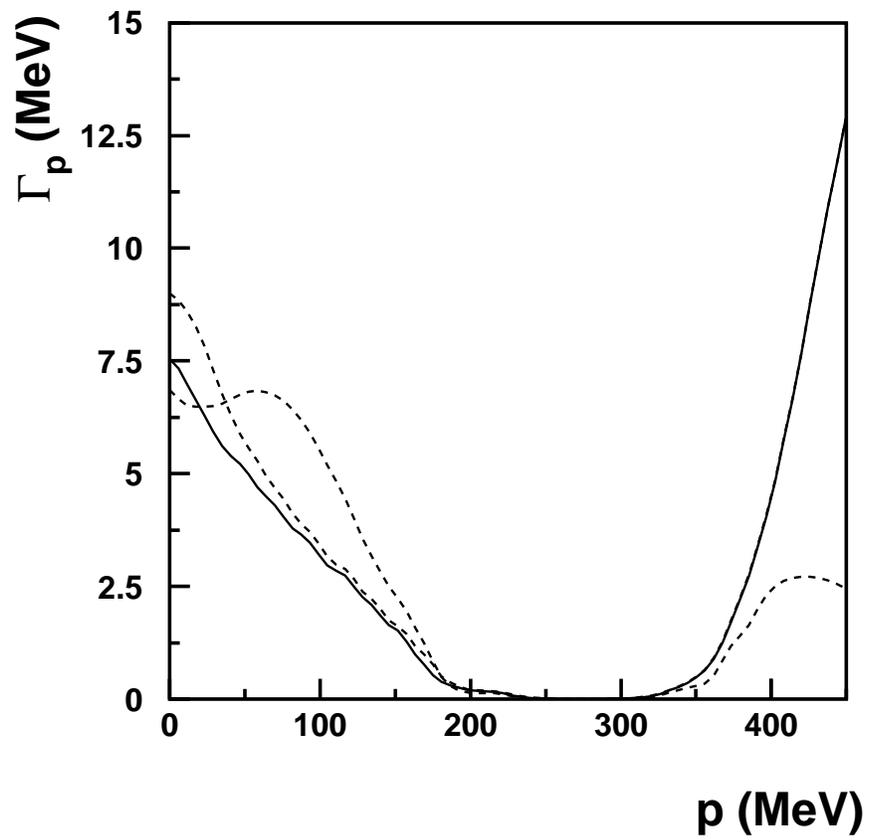}
\caption{The single-particle width at the quasi-particle pole of $A(p,\omega)$
(solid line) and at the quasi-particle poles of $A_s(p,\omega)$ (dashed lines).}
\label{widthfig}
\end{figure}

\begin{figure}

\centering
\includegraphics[width=0.9\textwidth]{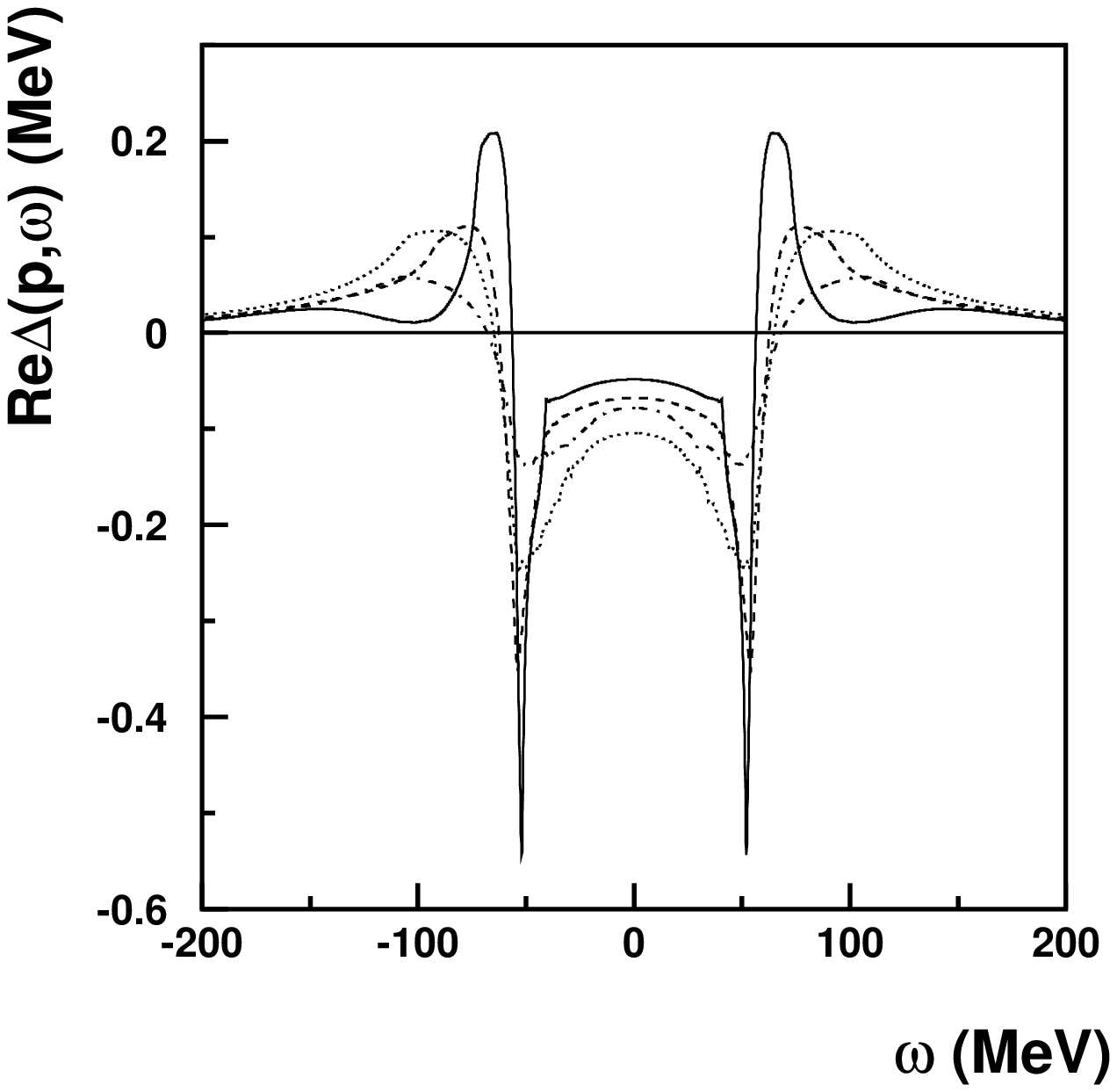}
\caption{The real part of the $T$-matrix contribution to the superfluid gap 
as function of energy for $p=0,116,233,350$MeV (solid, dashed, dotted and
dash-dotted lines).}
\label{deltrfig}
\end{figure}

\begin{figure}

\centering
\includegraphics[width=0.9\textwidth]{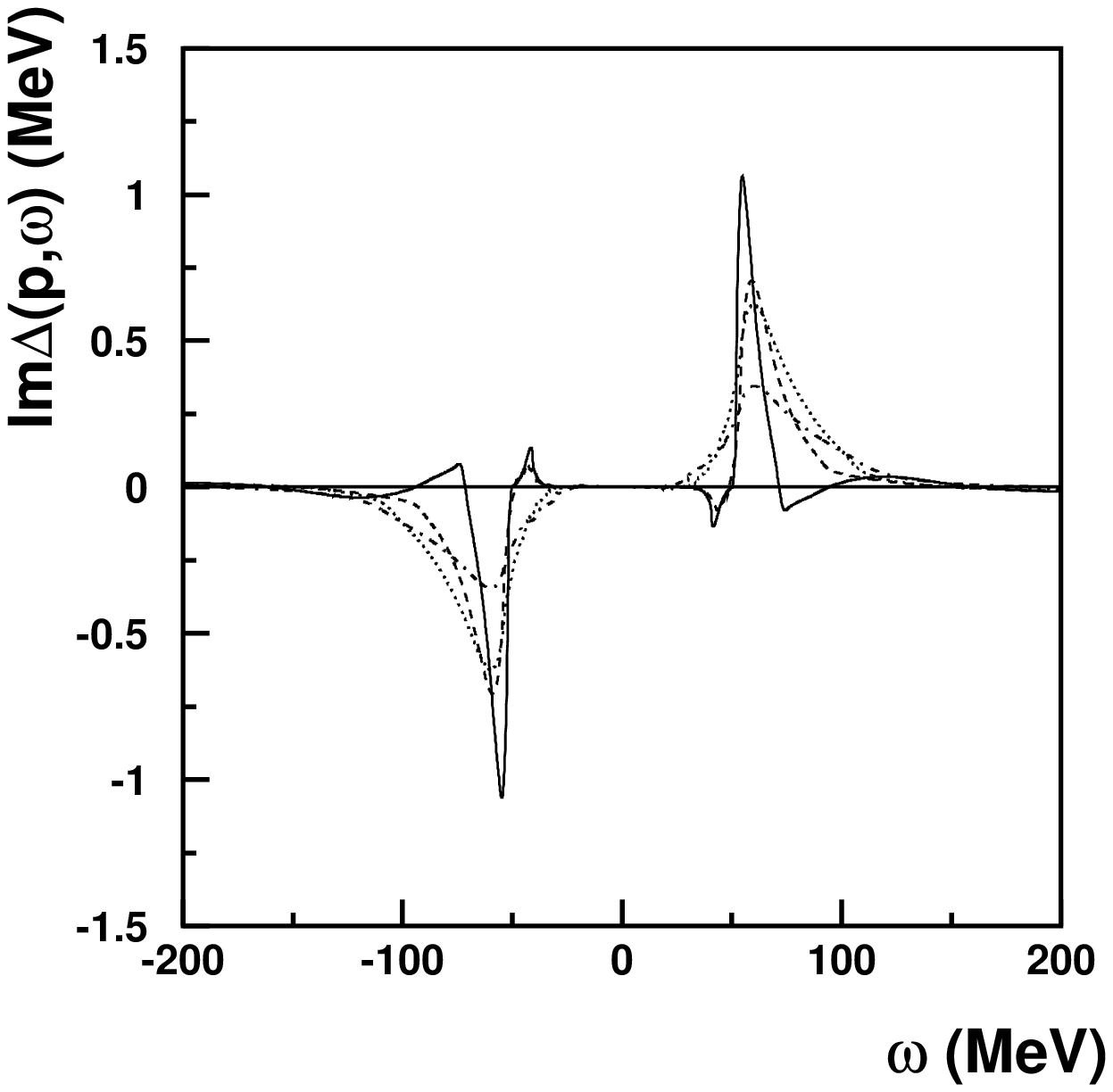}
\caption{The imaginary part of the $T$-matrix contribution to the superfluid gap 
as function of energy for $p=0,116,233,350$MeV (solid, dashed, dotted and
dash-dotted lines).}
\label{deltifig}
\end{figure}

\begin{figure}

\centering
\includegraphics[width=0.9\textwidth]{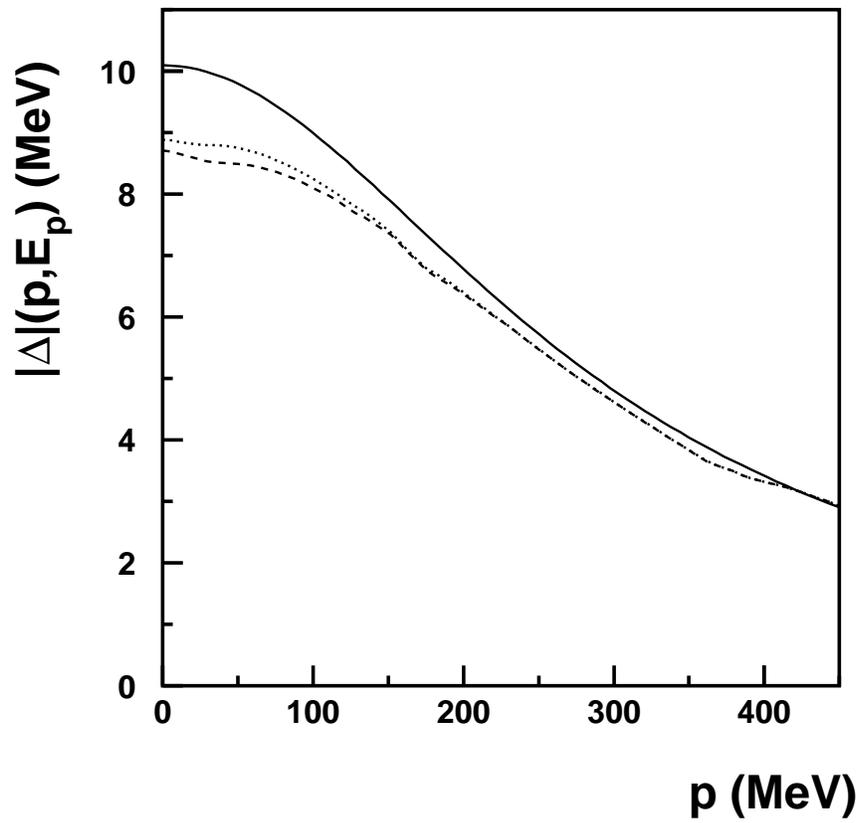}
\caption{The value of the superfluid gap at the quasi-particle pole of $\omega_p$
(dotted line) and at the pole $E_p$ (dashed line) compared to the BCS gap
(solid line).}
\label{deltfig}
\end{figure}

\clearpage

\begin{figure}

\centering
\includegraphics[width=0.8\textwidth]{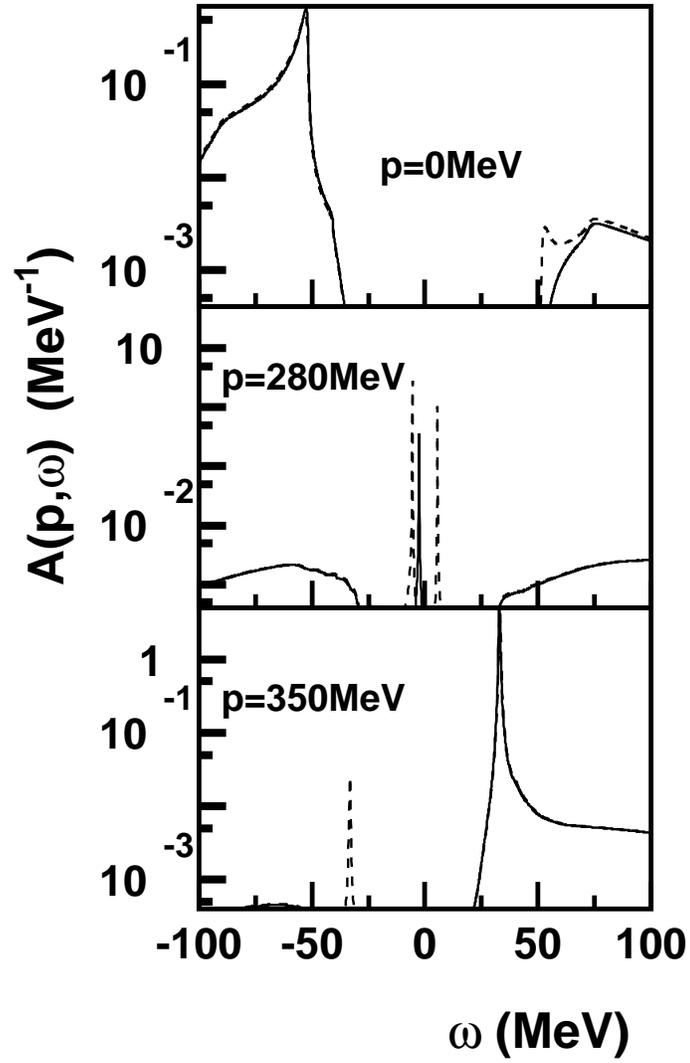}
\caption{The spectral function $A(p,\omega)$ including 
only the diagonal self-energy (solid lines) 
and the full spectral function $A_s(p,\omega)$
(dashed lines) as function of the energy. }
\label{specfig}
\end{figure}

\begin{figure}

\centering
\includegraphics[width=0.8\textwidth]{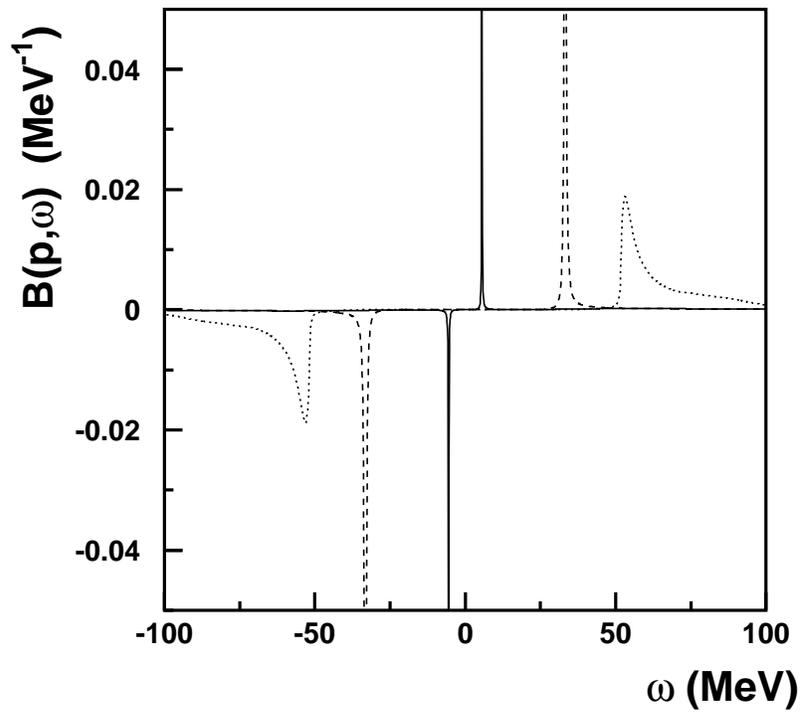}
\caption{The off-diagonal spectral function $B(p,\omega)$ 
for $p=0, 280$ and $350$MeV (dotted, solid and dashed lines respectively).}
\label{aspecfig}
\end{figure}

\end{document}